\documentclass[aps,prb,twocolumn,superscriptaddress]{revtex4-2}

\usepackage{graphicx}
\usepackage{dcolumn}
\usepackage{bm}

\usepackage{graphicx}
\usepackage{derivative}
\usepackage{dcolumn}
\usepackage{orcidlink}
\usepackage{bm}
\usepackage{hyperref}
\usepackage{braket}
\usepackage{amsmath}
\usepackage{xcolor}

\usepackage{amsfonts}
\usepackage[dvipsnames]{xcolor}

\begin{document}

\preprint{APS/123-QED}

\title{Study of localization properties under subsystem periodic driving}

\author{Arpita Goswami}
\author{Ranjan Modak}
\thanks{Contact author: ranjan@iittp.ac.in}
\author{Shaon Sahoo}
\thanks{Contact author: shaon@iittp.ac.in}
\affiliation{Department of Physics, Indian Institute of Technology Tirupati, India, 517619}

\date{\today}

\begin{abstract}
Periodic driving provides a powerful route for controlling localization and engineering effective Hamiltonians in quantum systems. A localized one-dimensional Aubry-Andr\'e chain can be driven into a fully delocalized phase when the entire system is driven below a critical frequency. Here, we ask whether such delocalization can be achieved by driving only a part of the system. We address this question using a two-leg ladder comprising an Aubry-Andr\'e chain coupled to a periodically driven clean tight-binding chain. Exact numerical calculations show that, starting from a localized static ladder, subsystem driving below a critical frequency cannot achieve complete delocalization. Instead, a mixed phase emerges, with localized and delocalized states coexisting and the fraction of delocalized states varying nonmonotonically with the driving frequency. This behavior can be understood using local effective Hamiltonians. Although these effective Hamiltonians are derived analytically in the high-frequency regime, they can nevertheless predict the localization transition occurring in the moderate-frequency regime. In fact, by quantifying the emergent nonlocality of the Floquet Hamiltonian, we show that these effective Hamiltonians provide a valid description up to a crossover frequency that is lower than the critical frequency associated with the localization transition. Our results demonstrate an alternative route to Floquet engineering, in which the transport properties of a quantum lattice system can be modified by driving only a subsystem.
\end{abstract}


\maketitle

\section{Introduction}

In low-dimensional non-interacting systems, arbitrarily weak disorder can suppress transport and exponentially localize single-particle states through Anderson localization ~\cite{AL_1, AL_2}. In presence of interaction, the disorder in the system can lead to many-body localization (MBL), characterized by the breakdown of thermalization and the Eigenstate Thermalization Hypothesis (ETH)~\cite{MBL1, MBL2, MBL3, MBL4, MBL_rev1, nandkishore2015many, sierant25, ETH1, ETH2, rigol2012alternatives, basko2006metal, SL_2, SL_1}.  
However, some recent studies claim that the rare thermal regions can destabilize the MBL phase through avalanche processes~\cite{MBL_breakdown1, MBL_breakdown2, vidmar.2020,sierant25}. 
In the three dimensional noninteracting systems, however, localized and extended states can coexist, separated by a mobility edge (ME)~\cite{pr_sc_2, anderson_3d}. 

On the other hand Aubry-Andr\'e (AA) model provides a paradigmatic example of a sharp, energy-independent localization transition in one dimension \cite{QP_1,QP_2,QP_3,QP_4,QP_5,QP_6,QP_7,QP_8,QP_9,QP_10,QP_11}, modifications of which, including potential perturbations and long-range hopping, can generate mobility edges~\cite{ME_1, ME_2, ME_3, ME_4, ME_5, ME_6}. These predictions have been experimentally explored in several platforms, including ultracold atoms~\cite{EXP_1,EXP_2,EXP_3,EXP_4,EXP_5,EXP_6,EXP_7,EXP_8,EXP_9}, photonic crystals~\cite{expt_1,expt_2,expt_3,expt_4}, electrical circuits~\cite{electric_circuit}, optical cavities, and superconducting circuits~\cite{sc_qubit_1,sc_qubit_2}. Quasiperiodic systems can also host a critical phase characterized by states that are extended yet nonergodic and exhibit scale-invariant properties, leading to unusual spectral statistics, fractal structures, and distinct dynamical behavior~\cite{critical_1,critical_2,critical_3,critical_4,nonergodic_1,spec_stat_1,spec_stat_3,spec_stat_4,spec_stat_5,frac_1,frac_2,frac_3,dyn_1,dyn_2}. 

In recent years, the idea of induced localization (or delocalization) in single-particle~\cite{Floquet_singleparticle1, Floquet_singleparticle2} and many-body~\cite{Floquet_manybody_1, floquet_manybody2} systems via tuning the hybridization of Wannier orbitals with time-periodic drives has attracted significant interest. Studies suggest that localization can be melted by a time-periodic drive that restores Wannier-function hybridization and induces delocalization.
In a similar spirit, localization can also be generated dynamically in a clean system by various controlled time-periodic drive protocols at resonant frequencies, by tuning a destructive interference of the extended wave functions. Such driving has enabled dynamical localization~\cite{dyn_loc_1, dyn_loc_2, dyn_loc_3}, coherent destruction of tunneling~\cite{tunn_1, tunn_2, tunn_3}, Floquet-engineered topological phases~\cite{Floquet_engineered_tp_1, Floquet_engineered_tp_2, Floquet_engineered_tp_3, Floquet_engineered_tp_4, Floquet_engineered_tp_5, Floquet_engineered_tp_6}, and quantum phase transitions~\cite{Quantum_phase_tran_1, Quantum_phase_tran_2}, with experimental realizations in ultracold atoms and photonic lattices~\cite{ultracold_1, ultracold_2, ultracold_3, ultracold_4, ultracold_5, ultracold_6}.



Floquet theory gives an effective Floquet Hamiltonian~\cite{eff_floquet_ham1} description of the long-time evolution of a periodically driven system, where the effective Floquet Hamiltonian depends on the drive amplitude and frequency. Particularly in quasiperiodic systems~\cite{ME_mosaic,gpd}, a periodic drive can further modify localization properties and induce a tunable mobility edge~\cite{ME_in_Flo_1, ME_in_Flo_2}, with the driving parameters serving as the tuning parameters. These studies, however, have primarily focused on driving the system as a whole.

\par This gives rise to a natural question: how are the localization properties of a composite system affected when only a part of the system, i.e., a subsystem, is periodically driven while the rest of the system is left undriven? To address this, we consider a ladder system consisting of an AA chain coupled to a periodically driven clean chain. 
The extensive study of the static counterpart of the whole system revealed a rich phase diagram~\cite{subsystem_localization} that naturally arises from hybridization between localized and extended states in the ladder geometry. 

In this study, we in particular ask whether periodic driving of the clean subsystem can induce a transition from a localized regime of the static system to a delocalized regime. We implement this subsystem driving by applying a square-wave drive exclusively to the clean leg. 
Using exact numerical calculations together with effective models derived analytically in the high-frequency regime, we find that, starting from a localized phase of the static ladder, there exists a critical driving frequency below which complete delocalization cannot be achieved through subsystem driving. Instead, the system enters a mixed phase in which localized and delocalized states coexist. Interestingly, within this mixed phase, the fraction of delocalized states varies nonmonotonically with the driving frequency.

The leading order effective Hamiltonians are derived using a rotating-frame transformation and a Floquet-van Vleck expansion, yielding effective descriptions in which the interleg hopping is frequency- and amplitude-dependent. 
We then compare these effective Hamiltonians with exact Floquet calculations and find that they accurately describe the high-frequency regime but gradually break down at lower frequencies. By comparing the effective and exact Floquet Hamiltonians, we show that this breakdown is associated with the emergence of drive-induced nonlocal couplings, which are absent in both the static system and the leading-order effective Hamiltonian. The exact Floquet Hamiltonian reveals a crossover frequency below which contributions from nonlocal hopping terms become appreciable. Interestingly, this crossover frequency lies below the critical frequency associated with the localization transition. This indicates that nonlocal couplings play a relatively minor role near the transition, and their effects can be effectively captured through a renormalization of the local (rung) coupling. Consequently, the effective Floquet Hamiltonians considered here, which retain only local couplings, are still capable of capturing the localization transition from the localized phase to the mixed phase.
 \par
\textcolor{black}{The remainder of this paper is organized as follows. In Sec.~\ref{mod_ham}, we introduce the two-leg ladder model and describe the subsystem-selective square-wave driving protocol. In Sec.~\ref{num}, we investigate the localization properties of the ladder, examining how the periodic drive modifies the Floquet eigenstates and the spatial structure of the Floquet Hamiltonian. We compare the localization crossover with the evolution of the spatial range of the Floquet couplings. In Sec.~\ref{ana}, we develop analytical effective-Hamiltonian descriptions using a rotating-frame transformation and a high-frequency van Vleck expansion. We then compare the localization properties predicted by these effective descriptions with the exact Floquet results by using the frequency-dependent fraction of delocalized states. Finally, Sec.~\ref{conc} summarizes our main findings and discusses the implications of subsystem-selective driving for controlling localization in coupled quantum systems.}

\begin{figure}
    \centering
    \includegraphics[width=0.5\textwidth]{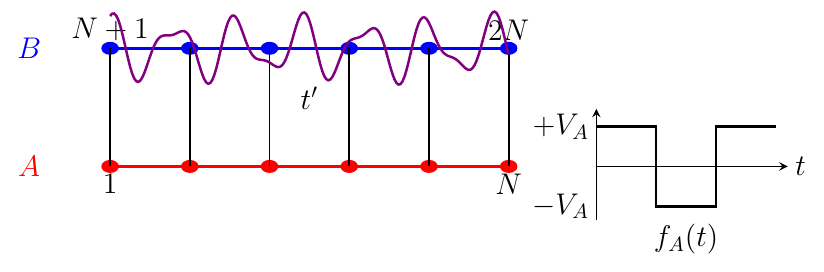}
    \caption{Schematic diagram of the model system, represented by Hamiltonian of Eq. \ref{ham_eqn}.}   \label{fig0}
\end{figure}

\section{Model and Driving Protocol} \label{mod_ham}

In this work, we investigate a ladder system comprising two coupled chains, referred to as the ``subsystem" ($A$) and the ``bath" ($B$), as illustrated schematically in Fig. \ref{fig0}. The subsystem $A$ is described by a clean tight-binding Hamiltonian ($H_A$ in Eq. \ref{ham_eqn}) subjected to a periodic square-pulse drive, while the bath $B$ is governed by an Aubry-Andr\'{e} (AA)-type Hamiltonian ($H_B$ in Eq. \ref{ham_eqn}). The two chains are coupled through inter-leg hopping of strength $t'$. The central question we address is whether periodic driving applied exclusively to subsystem $A$ can induce delocalization in the entire ladder when the corresponding undriven system is localized.

The full Hamiltonian of the periodically driven ladder system is given by
\begin{align}\label{ham_eqn}
\begin{split}
H(t)&=H_A+H_B+H_{AB}+f_A(t)N_A,~ \text{where}\\
H_A&=-t_A\sum_{n=1}^{N}
\left(a_n^\dagger a_{n+1}+\mathrm{H.c.}\right)\\
H_B& =-t_B\sum_{n=1}^{N}
\left(b_n^\dagger b_{n+1}+\mathrm{H.c.}\right) 
\\& +V_B\sum_{n=1}^{N}
\cos(2\pi\beta n)b_n^\dagger b_n~ \text{and}\\
H_{AB}&=-t'_{AB}\sum_{n=1}^{N}
\left(a_n^\dagger b_{n}+\mathrm{H.c.}\right).
\end{split}
\end{align}
We take $t_A=t_B=1$ and $\beta=\frac{\sqrt{5}-1}{2}$. For our numerical study, we take fixed values of $V_B$ and $t'_{AB}$ ($=t'$) so that the undriven system is deep in the localized phase. Here, $N_A=\sum_{n=1}^{N}a_n^\dagger a_n$, where $a_n^\dagger$ ($a_n$) denotes the creation (annihilation) operator at site $n$ of the clean leg A and $b_n^\dagger$ ($b_n$) are the same for leg B.

We apply a periodic square-wave potential to the clean leg,
\begin{equation} \label{squarewave}
f_A(t)=
\begin{cases}
+V_A, & 0\leq t<T/2,\\
-V_A, & T/2\leq t<T,
\end{cases}
 \qquad f_A(t+T)=f_A(t),
\end{equation}
where $V_A$ is the driving amplitude and $T=\frac{2\pi}{\omega}$ is the driving period, with $\omega$ denoting the driving frequency. Thus, the drive acts only on one leg of the ladder and periodically modulates its onsite energy relative to the AA leg.

For the time-periodic Hamiltonian $H(t+T)=H(t)$ in Eq.~\eqref{ham_eqn}, the stroboscopic evolution over one driving cycle is governed by the Floquet operator (with $\hbar=1$)
\begin{equation}
U_F(T)=\mathcal{T}\exp\!\left[-i\int_0^T H(t)\,dt\right],
\label{eq:UF_general}
\end{equation}
where $\mathcal{T}$ denotes time ordering. For the square-pulse drive in Eq.~\eqref{squarewave}, the Hamiltonian is piecewise constant, $H_\pm=H_0\pm V_A N_A$ with $H_0=H_A+H_B+H_{AB}$, so the time-ordered exponential reduces to a product of two exponentials:
\begin{equation}
U_F(T)=e^{-iH_-T/2}\,e^{-iH_+T/2}.
\label{eq:UF_square}
\end{equation}
The Floquet operator can be written in terms of a time-independent Floquet Hamiltonian $H_F$ as
\begin{equation}
U_F(T)=e^{-iH_F T},
\end{equation}
so the exact Floquet Hamiltonian is
\begin{equation}
H_F=\frac{i}{T}\log U_F(T).
\label{eq:eff_HF}
\end{equation}
Since $H(t)$ is quadratic in the fermionic operators, the dynamics can be studied in the single-particle sector, where $H_\pm$, and hence $U_F(T)$, are $2N\times 2N$ matrices in the site basis. At a given driving frequency $\omega$, we construct $U_F(T)$ from Eq.~\eqref{eq:UF_square} and obtain its spectrum by exact diagonalization. The quasienergies $\epsilon_n$ and the corresponding Floquet eigenstates $\ket{\psi_n}$, with $n=1,\dots,2N$, satisfy
\begin{equation}\label{eq:floquet_modes}
U_F(T)\ket{\psi_n}=e^{-i\epsilon_n T}\ket{\psi_n}.
\end{equation}
Using these Floquet eigenstates, we study their localization properties as a function of $\omega$ in the subsequent section. In addition, since $H_F$ obtained from Eq.~\eqref{eq:eff_HF} generally contains long-range hopping terms generated by the drive, we quantify its non-locality to understand how the effective couplings mediate delocalization in the ladder.

\begin{figure*}[t]
    \centering
    \includegraphics[width=0.96\textwidth]{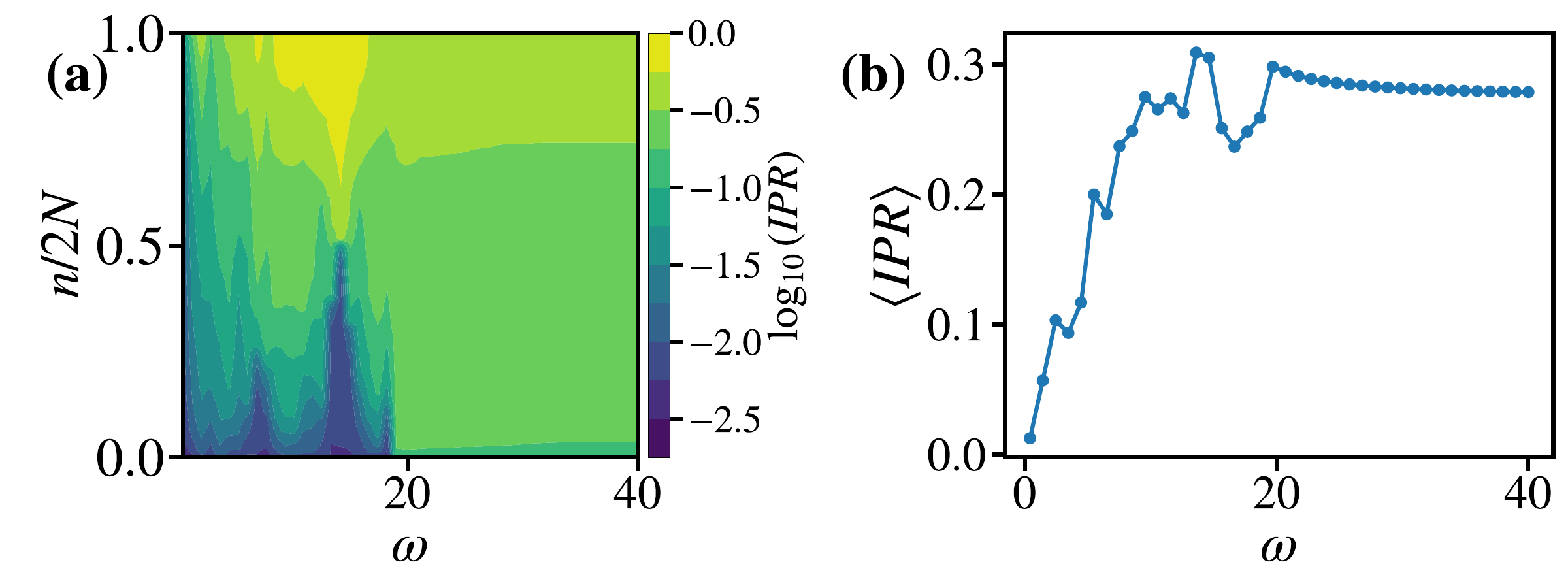}
    \caption{(a) Contour plot of sorted $\log{(IPR)}$ as a function of normalized state index and $\omega$, (b) plot of average IPR as a function of $\omega$ for $V_B=6$, $t'=10$, $V_A=20$ and total system size $2N=1000$. }
    \label{fig:cont1}
\end{figure*}

\section{Numerical results} \label{num}
In this section, we study localization and non-locality properties of the Floquet Hamiltonian $H_F$. We choose $t'=10$ and $V_B=6$ to make sure that the undriven system is in the deep inside the localized phase (see Appendix \ref{static_loc}). We also choose $V_A=20$ for our main calculations, and the effect of variation in $V_A$ is shown in Appendix~\ref{app_3}. 

\subsection{Localization property of the driven ladder}

To characterize the localization properties of the ladder system, we calculate the state-averaged inverse participation ratio (IPR) and normalized participation ratio (NPR) over the complete single-particle spectrum of $H_F$. For the $k-$th normalized eigenstate $\ket{\psi_k}$, the IPR is defined as \cite{ipr_1, ipr_2, ipr_3}
\begin{equation} \label{stat_IPR}
IPR^{(k)}=\sum_{j=1}^{2N}
\left|\langle j|\psi_k\rangle\right|^4 .
\end{equation}
The IPR quantifies the spatial localization of an eigenstate. For a state localized over a finite number of sites, $IPR^{(k)}$ remains finite in the thermodynamic limit, whereas for an extended state spread over the (2N) sites of the ladder, $IPR^{(k)}\sim 1/(2N)$ and therefore vanishes as $N\rightarrow\infty$.

The state-averaged IPR is given by
\begin{equation}
\langle IPR\rangle=
\frac{1}{2N}\sum_{k=1}^{2N} IPR^{(k)} .
\end{equation}
Consequently, $\langle IPR\rangle$ approaches a finite value for a predominantly localized spectrum, while it vanishes in the thermodynamic limit for a fully extended spectrum.

As a complementary diagnostic, we consider the normalized participation ratio (NPR),
\begin{equation}
NPR^{(k)}=
\frac{PR^{(k)}}{2N},
\end{equation}
where
\begin{equation}
PR^{(k)}=\frac{1}{IPR^{(k)}} .
\end{equation}
The corresponding state-averaged NPR is
\begin{equation}
\langle NPR\rangle=
\frac{1}{2N}\sum_{k=1}^{2N}NPR^{(k)}
=\frac{\langle PR\rangle}{2N}.
\end{equation}
For an extended state, $PR^{(k)}\propto N$, giving a finite NPR in the thermodynamic limit. In contrast, for a localized state, the participation ratio remains finite and hence $NPR^{(k)}\rightarrow0$ as $N\rightarrow\infty$. Critical or multifractal states likewise generally yield a vanishing NPR in this limit, although their finite-size scaling can differ from that of exponentially localized states.
\par
Since our primary objective is to investigate how a selective periodic drive applied to leg $A$ modifies the localization properties of the ladder, we first choose a strongly coupled regime $t'=10$, where the undriven system is well within the localized region. This provides a convenient localized reference point from which the drive's effect can be systematically investigated. The plots for the static Hamiltonian can be seen in Appendix~\ref{static_loc} (see Fig.~\ref{fig:ipr_npr}), which show the state-averaged IPR and NPR of the undriven ladder as a function of $V_B$ for $t'=10$. The $\langle NPR \rangle $ decreases as $V_B$ increases and becomes negligible beyond $V_B\simeq5$ ($\sim\mathcal O(1/2N)$), indicating that the spectrum is predominantly localized in this parameter regime. We therefore choose $V_B=6$ and $t'=10$ for the subsequent analysis. This parameter set places the undriven ladder well inside the localized regime and provides a controlled starting point for investigating the effect of the selective square-wave drive applied to leg $A$.

Having obtained the Floquet eigenstates for a given $\omega$ value from Eq. \ref{eq:floquet_modes}, we first analyze the IPR of the normalized Floquet eigenstate $\ket{\psi_n}$ using the formula in Eq.~\ref{stat_IPR}. We fix $V_B =6$, $t'=10$, $V_A=20$ and $2N =1000$. By varying the $\omega$ value, we plot the contour of the sorted IPR (sorted from lowest to highest value or in ascending order) as a function of normalized state index on the y-axis and $\omega$ on the x-axis. The result can be found in Fig.~\ref{fig:cont1}(a). We can see that, unlike in the case of an AA model, where the drive takes the system from a completely localized to a completely delocalized phase~\cite{ME_in_Flo_1}, here we obtain coexistence of localized and delocalized states, thereby forming a mixed-phase structure as $\omega$ decreases. However, above a critical value $\omega_c\simeq17$, we see a sharp change in localization. Below this frequency, the system exhibits a finite fraction of delocalized states, which changes non-monotonically with a change in $\omega$, and the system becomes localized monotonically above that frequency. We also plot the average IPR as a function of $\omega$ in Fig.~\ref{fig:cont1}(b), which shows a non-monotonic behavior below $\omega_c$, persisting with the presence of mixed phase, which generally occurs due to the non-monotonicity arising from the bond-localization.

\begin{figure*}[t]
\centering
\includegraphics[width=0.95\textwidth]{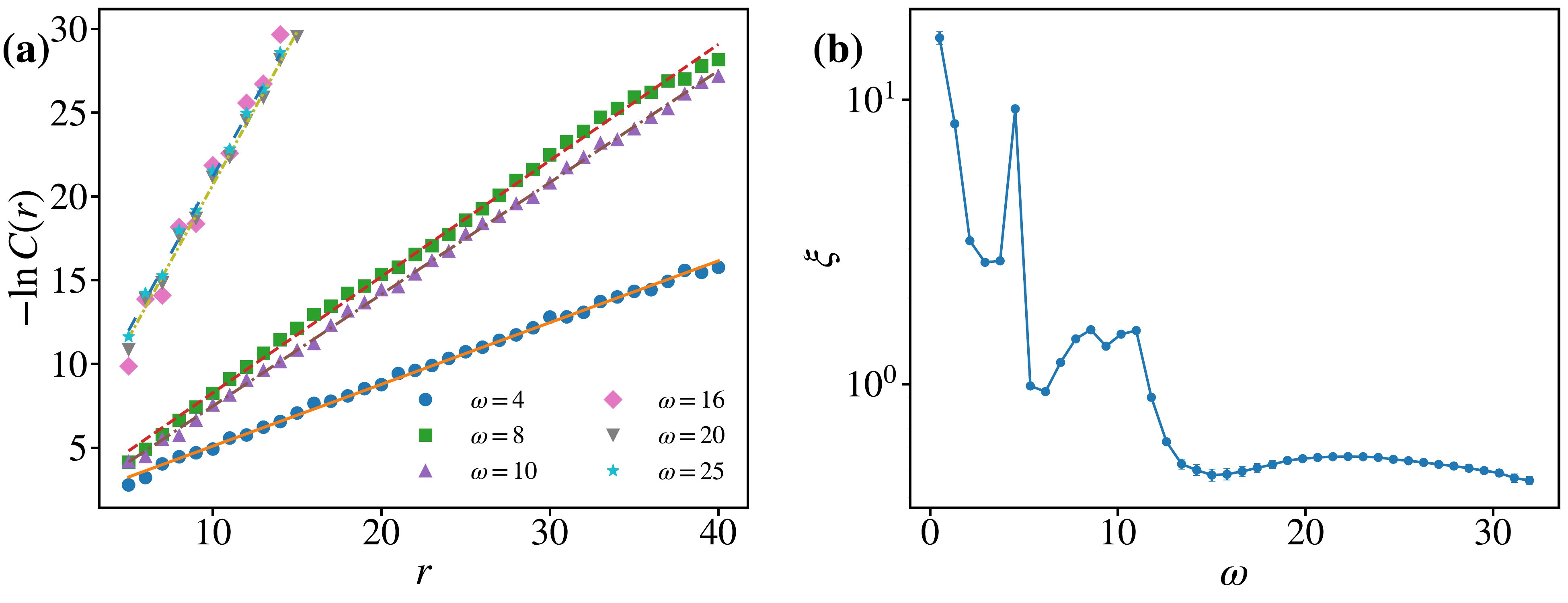}
 \caption{(a) Plot of $-\ln C(r)$ as a function of $r$ for different chosen values of $\omega$, (b) Plot of $\xi$ as a function of $\omega$ at $V_B=6$, $t'=10$ and $V_A=20$.}
    \label{fig:lnC_vs_r}
\end{figure*}

\subsection{Driving induced nonlocality in $H_F$}

To gain deeper insight into the localization properties of the driven system, we next examine the spatial structure of the exact Floquet Hamiltonian. Specifically, we investigate how the spatial distribution of its matrix elements evolves with the driving frequency and identify a crossover frequency below which nonlocal couplings become significant and can no longer be neglected.

\textcolor{black}{To quantify the drive-induced nonlocality across the ladder, we introduce the distance-resolved inter-leg coupling amplitude:}

\textcolor{black}{\begin{equation} \label{eq:Cr_AB}
C(r) = \frac{1}{N_r} \sum_{i=1}^{N - r} \left| (H_F)_{i, \, N+r+i} \right|,
\end{equation}}

\textcolor{black}{where $N_r = N - r$ denotes the number of inter-leg matrix elements included in the average at spatial separation $r$ ($r$ can take values $0, \dots N-1$), and $(H_F)_{i,\, N+r+i}$ connects site $i$ on leg $A$ to site $N+r+i$ on leg $B$. This quantity characterizes the average magnitude of the effective cross-ladder couplings as a function of spatial separation. At $r = 0$, $C(0)$ gives the average magnitude of the local rung couplings connecting site $i$ on leg $A$ to site $N+i$ on leg $B$ (corresponds to rung-coupling in the static Hamiltonian). For $r > 0$, $C(r)$ measures the drive-induced long-range diagonal hoppings generated between the two chains.} 
To characterize the spatial structure of the Floquet Hamiltonian, we examine the distribution of its matrix elements for different driving frequencies. For comparison, we first consider the static Hamiltonian, whose matrix-element distribution reflects its short-range coupling structure. The corresponding matrix-element plots for the static (undriven) case and for $\omega=40,20, 10$ and $5$ of $H_F$ are shown in Appendix Fig.~\ref{fig:mat_ele}. Comparing these plots allows us to track the development of longer-range matrix elements as the driving frequency is reduced.

\par At high driving frequencies, the matrix elements of the Floquet Hamiltonian decay rapidly with distance, consistent with a predominantly local effective description. As the frequency decreases, the distance dependence of $C(r)$ changes, indicating that longer-range couplings become increasingly relevant. To quantify this behavior, we first examine whether the coupling profile can be described by an exponential decay of the following form:
\textcolor{black}
{\begin{equation}
C(r) \sim C(0)~ e^{-r/\xi}
\end{equation}}

\textcolor{black}{Here, $C(0)$ is a multiplicative constant that can be obtained from Eq.~\ref{eq:Cr_AB} by putting $r=0$, and $\xi$ denotes the decay length of the coupling amplitude.} 

Figure~\ref{fig:lnC_vs_r}(a) shows the distance dependence of the inter-leg coupling amplitude for representative driving frequencies, while Fig.~\ref{fig:lnC_vs_r}(b) presents the extracted decay length $\xi$ as a function of frequency. The results indicate that the spatial range of the cross-leg Floquet couplings depends strongly on the driving frequency. In particular, the growth of the decay length as the frequency is lowered signals the increasing importance of longer-range inter-leg matrix elements in the low-frequency regime.

From Fig.~\ref{fig:lnC_vs_r}(b), we roughly identify a crossover frequency $\omega_{\mathrm{cr}}\simeq 15$, below which the inter-leg nonlocality becomes significant. For the same parameters, Fig.~\ref{fig:cont1} shows that the localization transition occurs at $\omega_c\simeq 17>\omega_{\mathrm{cr}}$, implying that the inter-leg nonlocality remains weak even at the transition point. This observation motivates the analytical study presented in the subsequent section, where we derive an effective local Hamiltonian in the high-frequency regime and show that it captures the localization transition observed in the exact Floquet dynamics.

\section{Analytical results} \label{ana}
In this section, we provide an analytical understanding of our numerical results. We employ two complementary techniques, (i) a rotating-frame transformation and (ii) the van Vleck high-frequency expansion, to derive effective time-independent Hamiltonians. These are expected to describe the exact Floquet Hamiltonian accurately in the high-frequency regime. In the subsequent section, we compare our exact numerical results with those obtained from these effective Hamiltonians.

\subsection{Rotating-frame transformation}
We first discuss the rotating-frame transformation, in which the explicit time dependence of the driving term is eliminated by the unitary transformation
\begin{equation}
U(t)=\exp\!\left[-i\phi(t)N_A\right], \nonumber
\label{Rotation}
\end{equation}
with the rotated state defined as
\begin{equation}
|\tilde{\psi}(t)\rangle=U^\dagger(t)\,|\psi(t)\rangle .\nonumber
\end{equation}
Here, $\phi(t)$ can be any arbitrary real function. 
The Hamiltonian governing the dynamics in the rotating frame can then be read as, 
\begin{equation}
\tilde H(t)=U^\dagger(t)H(t)U(t)
-iU^\dagger(t)\frac{dU(t)}{dt}.
\label{rot_H}
\end{equation}
Using
\begin{equation}
\frac{dU(t)}{dt}=-i\dot{\phi}(t)\,N_A\,U(t),\nonumber
\end{equation}
Eq.~\eqref{rot_H} becomes
\begin{equation}
\tilde H(t)=U^\dagger(t)H_0U(t)+\left[f(t)-\dot{\phi}(t)\right]N_A,
\end{equation}
where
\begin{equation}
H_0=H_A+H_B+H_{AB}.\nonumber
\end{equation}
Choosing
\begin{equation}
\dot{\phi}(t)=f(t)\nonumber 
\end{equation}
eliminates the explicit driving term entirely, so that
\begin{equation}
\tilde H(t)=U^\dagger(t)H_0U(t).
\label{transformed_H}
\end{equation}

Since $N_A$ commutes with all operators acting exclusively on leg $B$, $H_B$ is unchanged by this transformation. The operators on leg $A$, by contrast, acquire time-dependent phases,
\begin{equation}
U^\dagger(t)\,a_n\,U(t)=e^{-i\phi(t)}a_n,\qquad
U^\dagger(t)\,a_n^\dagger\,U(t)=e^{i\phi(t)}a_n^\dagger . \nonumber
\end{equation}

The intraleg hopping on the clean chain is therefore also invariant, because neighboring sites acquire the same phase. The transformed Hamiltonian thus takes the form

\begin{equation}\label{rotated_H}
\begin{split}
\tilde H(t) = &
-t_A\sum_n \left(a_n^\dagger a_{n+1}+\mathrm{H.c.}\right)
-t_B\sum_n \left(b_n^\dagger b_{n+1}+\mathrm{H.c.}\right)\\
&+V_B\sum_n \cos(2\pi\beta n)b_n^\dagger b_n\\
&-t'\sum_n \left( e^{-i\phi(t)}a_n^\dagger b_n
+ e^{i\phi(t)}b_n^\dagger a_n \right).
\end{split}
\end{equation}

The periodic on-site modulation of one leg is thus mapped onto a time-dependent Peierls phase~\cite{ladder_peierls, peierls} on the interleg hopping. In this representation, the drive acts directly on the hybridization between the two legs~\cite{subsystem_localization}.

For the square-wave protocol of Eq.~\eqref{squarewave}, integrating $\dot{\phi}(t)=f(t)$ gives
\begin{equation}
\phi(t)=
\begin{cases}
V_At, & 0<t<T/2,\\
V_A(T-t), & T/2<t<T.
\end{cases}
\label{phase}
\end{equation}
The time dependence of the rung coupling can consequently be expanded in Fourier harmonics as
\begin{equation} \label{fourier_exp}
e^{i\phi(t)}= \sum_{m=-\infty}^{\infty}
c_m e^{-im\omega t},
\end{equation}
where
\begin{equation} \label{fourier_coeff}
c_m= \frac{1}{T} \int_0^T dt e^{i\phi(t)}e^{im\omega t}.
\end{equation}

\paragraph*{Leading-order Floquet Hamiltonian:}
The dynamics of a driven system can be systematically analyzed with a Floquet theory~\cite{Floqute_theory_1, Floqute_theory_2}, where the stroboscopic evolution over one driving period is generated by an effective time-independent Floquet Hamiltonian $ H_F$.
Due to the time ordering operation in Eq.~\ref{eq:UF_general}, the exact expression for $H_F$ is difficult to compute. However, one can employ high-frequency expansions to obtain the effective Hamiltonian when the energy scale associated with the driving frequency dominates. The effective Floquet Hamiltonian is given by $ H_F =  H_F^{(0)}+ H_F^{(1)}+ H_F^{(2)} + \dots$, where $ H_F^{(n)}$ scales as $\omega^{-n}$. At sufficiently high driving frequencies, the leading contribution to our transformed Floquet Hamiltonian ($\tilde H(t)$) is given by the time average of the leading order component of the rotated Hamiltonian over one driving period,
\begin{equation}
 H_F \simeq H_F^{(0)}=\frac{1}{T} \int_0^T dt\,\tilde H(t).
\label{zeroth_Floquet}
\end{equation} 
Using Eq.~\ref{rotated_H} and Eq.~\ref{fourier_exp} which can be explicitly written as follows:
\[ H_F^{(0)}=\frac{1}{T} \int_0^T dt\,c_0 \,+ \frac{1}{T} \int_0^T dt\,\sum_{m \ne 0}
c_m e^{-im\omega t}.\]
Now one can interchange the order of the summation over $m$ and the integration over $t$ and write:
\[ H_F^{(0)}=\frac{1}{T} \int_0^T dt\,c_0 \,+ \frac{1}{T} \sum_{m \ne 0} c_m \int_0^T dt \, e^{-im\omega t}\]

\textcolor{black}{Only the zeroth Fourier component of the time-dependent rung phase contributes, as for $m \ne 0$, $\int_0^T dt \, e^{-im\omega t}=0$ for one period }. For the square-wave protocol, the $c_0$ component is
\begin{equation}
c_0= e^{iV_AT/4}
\,
\mathrm{sinc}\left(\frac{V_AT}{4}\right),
\qquad
\mathrm{sinc}(x)=\frac{\sin x}{x}.
\label{c0}
\end{equation}
Consequently, the leading-order Floquet Hamiltonian can be written as
\begin{equation} \label{HF0}
\begin{split}
 H_F^{(0)}=&
-t_A\sum_n \left(a_n^\dagger a_{n+1}+\mathrm{H.c.}\right)
-t_B\sum_n\left(b_n^\dagger b_{n+1}+\mathrm{H.c.}\right)\\
&+V_B\sum_n \cos(2\pi\beta n)b_n^\dagger b_n\\
&-t'_{\mathrm{eff}}
\sum_n\left(e^{-iV_AT/4}a_n^\dagger b_n +\mathrm{H.c.}\right),
\end{split}
\end{equation}
where the effective rung hopping is
\begin{equation} \label{teff}
t'_{\mathrm{eff}} = t'\mathrm{sinc}\left(\frac{V_AT}{4}\right).
\end{equation}

Thus, at leading order, the periodic drive preserves the local structure of the ladder and simply renormalizes the inter-leg rung-hopping amplitude, while also generating a uniform phase on the rung coupling. In this high-frequency regime, the driven system can therefore be described by an effective static ladder with a renormalized rung hopping. As shown below, this local description accurately reproduces the exact Floquet results at sufficiently large $\omega$, but breaks down at lower frequencies where higher-order Floquet processes become important.

The phases $e^{-iV_AT/4}$ can be eliminated by performing a local gauge transformation of the fermionic operators,
\begin{equation}
a_n = e^{-i\tilde{\theta}_n} c_n .\nonumber 
\end{equation}
where $\tilde{\theta}_n = V_AT/4$, which makes $e^{i\tilde{\theta}_n}$ a global and uniform phase across all the sites. Substituting this into the Hamiltonian $ H_F^{(0)}$ the Hamiltonian (Eq.~\eqref{HF0}) reduces to

\begin{equation} \label{HF0_final}
\begin{split}
 H_F^{(0)}=&
-t_A\sum_n \left(c_n^\dagger c_{n+1}+\mathrm{H.c.}\right)
-t_B\sum_n \left(b_n^\dagger b_{n+1}+\mathrm{H.c.}\right)\\
&+V_B\sum_n \cos(2\pi\beta n)b_n^\dagger b_n\\
&-t'_{\mathrm{eff}}
\sum_n\left(c_n^\dagger b_n+\mathrm{H.c.}\right).
\end{split}
\end{equation}

However, the higher-order terms (e.g., ${H}_F^{(1)}$ and ${H}_F^{(2)}$) generate long-range inter-leg hoppings in the effective Hamiltonian $H_F$. Nevertheless, at high driving frequencies $\omega$, these contributions are expected to remain small. To quantify this, we examine the coefficients $c_m$ associated with the different harmonic terms. 

From Eq.~\ref{c0}, we have 
\[c_0= e^{iV_AT/4}
\,
\mathrm{sinc}\left(\frac{V_AT}{4}\right),\]

The corresponding Fourier weight is given as follows:
\begin{equation}
|c_0|^2 = \text{sinc}^2\left(\frac{\pi V_A}{2\omega}\right) = \left[ \frac{\sin\left(\frac{\pi V_A}{2\omega}\right)}{\frac{\pi V_A}{2\omega}} \right]^2.
\end{equation}

Using Parseval's identity, we get:
\begin{equation}
\sum_{m=-\infty}^{\infty} |c_m|^2 = \frac{1}{T} \int_0^T \left| e^{i\phi(t)} \right|^2 dt = 1.
\end{equation}

At high-frequency limit ($\omega \gg V_A$), by Taylor expansion of $|c_0|^2$ around $x = \frac{\pi V_A}{2\omega} \to 0$ we get:
\begin{equation}
\text{sinc}(x) = 1 - \frac{x^2}{6} + \mathcal{O}(x^4).
\end{equation}
Thus,
\begin{equation}
|c_0|^2 = 1 - \frac{x^2}{3} + \mathcal{O}(x^4) = 1 - \frac{\pi^2 V_A^2}{12\omega^2} + \mathcal{O}\left(\frac{1}{\omega^4}\right).
\end{equation}

This gives us the following 

\begin{equation}
1 - |c_0|^2 = \frac{\pi^2 V_A^2}{12\omega^2} + \mathcal{O}\left(\frac{1}{\omega^4}\right) \xrightarrow{\omega \to \infty} 0.
\end{equation}
This result justifies retaining only $c_0$ as the leading-order contribution at large $\omega$, even when the higher-order terms ${H}_F^{(k)}$ with $k>0$ are taken into account.

\subsection{van Vleck high-frequency expansion}

We complement the rotating-frame analysis with a systematic high-frequency expansion. Writing
\begin{equation}
H(t)= H_0+\sum_{m=-\infty}^{\infty}H_m e^{im\omega t},
\end{equation}
where the Fourier components of the square-wave drive are
\begin{equation}
H_m=f_m N_A,
\end{equation}
where
\begin{equation}
f_m=
\begin{cases}
\dfrac{2V_A}{i\pi m}, & m~\text{odd},\\[6pt]
0, & m~\text{even}.
\end{cases}
\end{equation}

Taking the terms up to the order of $\mathcal O(1/\omega^2)$, the resulting effective static Hamiltonian can be written as:
\begin{equation}
    H_{\rm vV} = H_{0} +H^{(2)}
\end{equation}
Details of the calculations are presented in Appendix~\ref{app:van_vleck}. Substituting the values of $H_{0}$ and $H^{(2)}$ we get the following final expressions:
\begin{equation} \label{eq:H_vV}
    H_{\rm vV} = H_A +H_B -t'\left(1-\frac{\pi^2 V_A^2}{24 \omega^2}\right)\sum_{n}\left(a^{\dagger}_nb_n +b^{\dagger}_n a_n\right)
\end{equation}

\begin{figure*}[t]
    \centering
    \includegraphics[width=0.95\textwidth]{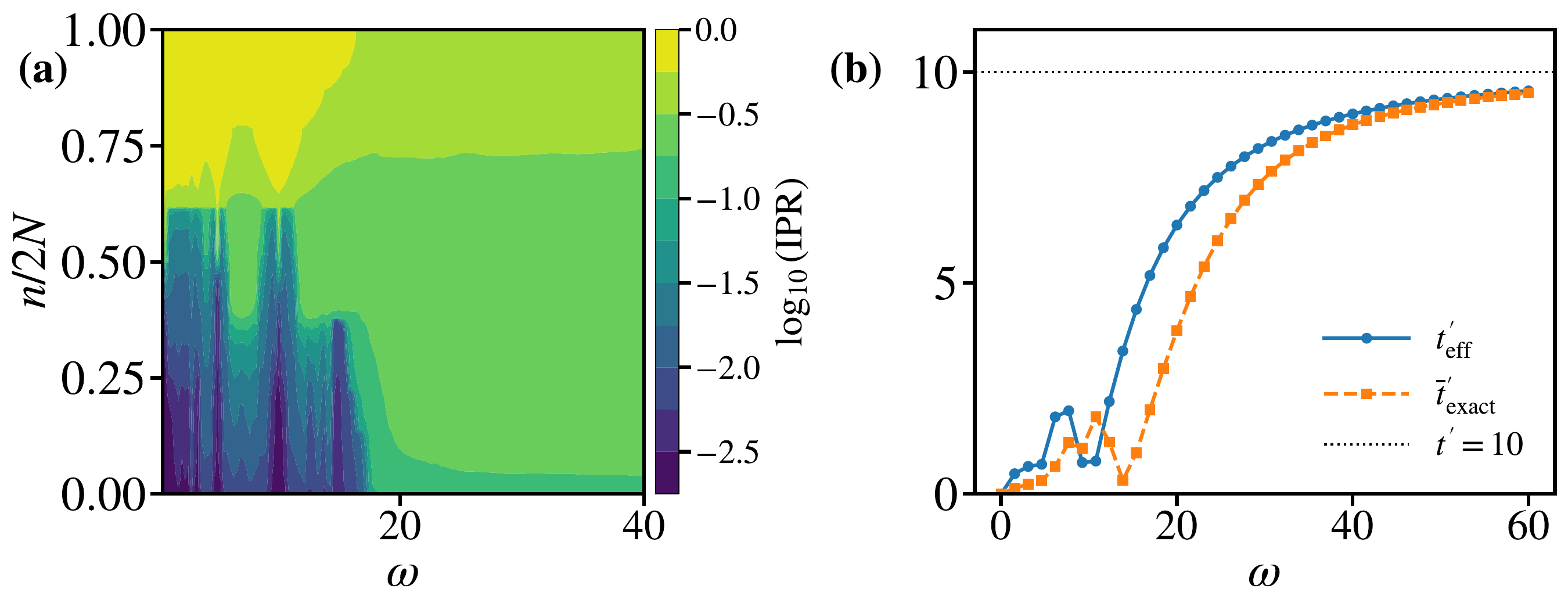}
    \caption{(a) Contour plot of sorted $\log{IPR}$ as a function of normalized state index and $\omega$ obtained from the effective Hamiltonian in Eq.~\ref{HF0_final}, (b) comparison between the analytically predicted effective rung hopping $t'_{\mathrm{eff}}$ and the average nearest-rung matrix element of the exact Floquet Hamiltonian,$\overline{|t'|}_{\mathrm{exact}}$ for $V_B=6$, $t'=10$, $V_A=20$ and $t_A=t_B =1$}
    \label{fig:cont2}
\end{figure*}

The second-order van Vleck expansion provides a local effective description of the driven ladder in terms of a renormalized rung coupling. From Eq.~(\ref{eq:H_vV}), the effective rung amplitude is given by
\begin{equation}  \label{eq:teff_vv}
    t'_{\rm eff}(\omega)=
    t'\left(1-\frac{\pi^2 V_A^2}{24\omega^2}\right).
\end{equation}
Thus, within the order retained in the high-frequency expansion, the effect of the drive can be interpreted as a frequency-dependent renormalization of the bare interleg coupling $t'$. In this description, the effective Hamiltonian retains exactly the same spatial connectivity as the undriven ladder; only the strength of the rung coupling is modified.

Equation~(\ref{eq:teff_vv}) also provides a natural way to manifest the frequency dependence of the problem. In the high-frequency limit $ \omega \to \infty$, the drive-induced correction vanishes and $t'_{\rm eff} \to  t'$. Consequently, the effective Floquet Hamiltonian approaches the undriven static ladder, $H_{\rm vV}\to H_0,$ as expected. This limit, therefore, provides the static reference regime in which the local Floquet description is recovered. Conversely, Eq.~(\ref{eq:teff_vv}) predicts a nominal vanishing of the effective rung coupling when $t'_{\rm eff}=0,$ which occurs at the characteristic frequency
\begin{equation}\label{eq:omega0}
\omega_0 = \frac{\pi V_A}{\sqrt{24}} = \frac{\pi V_A}{2\sqrt{6}}.
\end{equation}
At this frequency, the second-order local van Vleck Hamiltonian reduces to $H_{\rm vV}(\omega_0) = H_A+H_B,$ and hence predicts dynamical decoupling of the two legs at this level of approximation.

Equation~(\ref{eq:teff_vv}) is obtained from a truncated high-frequency expansion and is controlled only when the relevant dimensionless ratios, in particular $A/\omega$, remain sufficiently small. As $\omega$ is reduced toward $\omega_0$, higher-order terms in the Floquet expansion become increasingly important. The vanishing of $t'_{\rm eff}$ therefore marks the point at which the local second-order description predicts the strongest suppression of the rung coupling, rather than an exact decoupling of the full driven system. This analysis is particularly useful to choose a correct set of parameters, especially the $\omega$ values for which an effective description can give a valid prediction for the actual Floquet Hamiltonian.

\begin{figure*}
    \centering
    \includegraphics[width=1\textwidth]{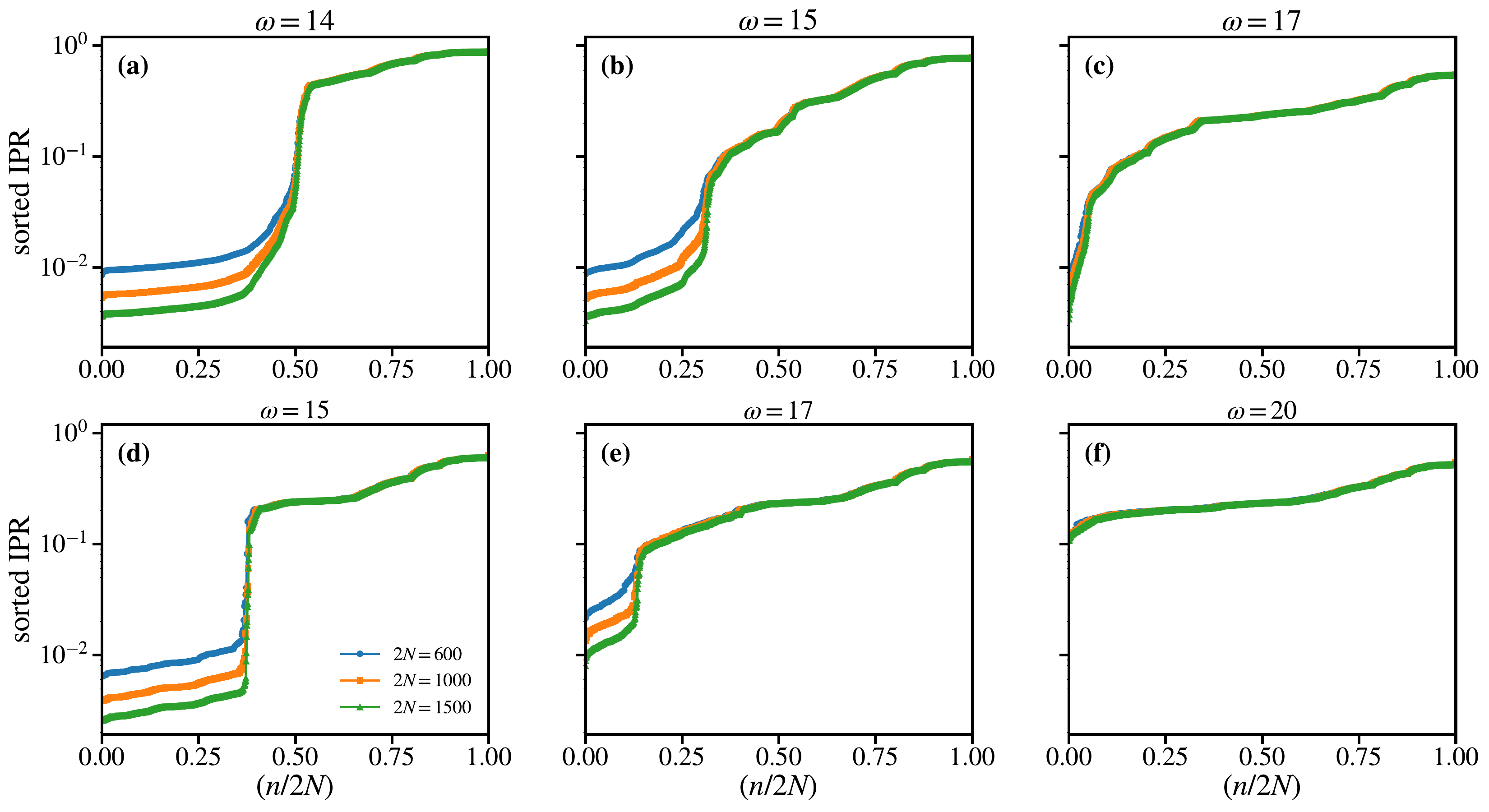}
    \caption{Sorted inverse participation ratio spectra of Floquet ladder eigenstates for different driving frequencies $\omega$. Panels (a–c) correspond to $\omega=14,~15$ and $17$ obtained from the exact Floquet Hamiltonian, panels (d–f) correspond to $\omega=15,~17$ and $20$ obtained from the effective analytical Hamiltonian in Eq.~\ref{HF0_final}, respectively with $V_B=6$ and $t'=10$ and $V_A=20$. Each curve compares system sizes $N=600, 1000, 1500$.}
    \label{sorted_iprs}
\end{figure*}

\begin{figure}[t]
    \centering
    \includegraphics[width=0.45\textwidth]{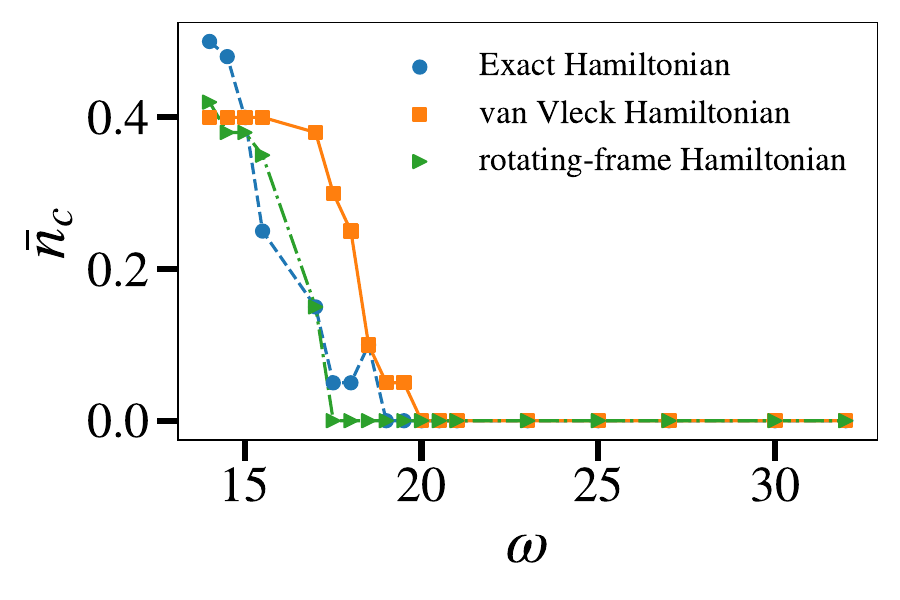}
    \caption{Fraction of delocalized Floquet eigenstates as a function of driving frequency.}
    \label{fig:n_c_vs_omega}
\end{figure}

\subsection{Comparison between  the effective Hamiltonian  and exact $H_F$}

We now investigate the localization properties of the effective Hamiltonian obtained in Eq.~\ref{HF0_final}, using the same parameters as in the exact Floquet calculations: $V_B=6$, $t'=10$, and $V_A=20$. Figure~\ref{fig:cont2}(a) shows the contour plot of the sorted inverse participation ratio (IPR) as a function of the normalized state index and the driving frequency $\omega$. The effective Hamiltonian reproduces the principal features of the exact Floquet results in Fig.~\ref{fig:cont1}(a), including the transition from the mixed to localized regime. Moreover, the localization transition frequency $\omega_c$ obtained from the effective model agrees closely with that extracted from the exact Floquet Hamiltonian $H_F$ (both cases $\omega_c\simeq 17$).

Although the effective Hamiltonian captures the localization transition, the details of the phase diagram still differ, particularly in the low-frequency regime, as is evident from a comparison of Figs.~\ref{fig:cont1}(a) and \ref{fig:cont2}(a). 
To understand these deviations, we compare the effective inter-leg hopping,
\begin{equation}
t'_{\mathrm{eff}}=t'\left|
\operatorname{sinc}\left(\frac{\pi V_A}{2\omega}\right)
\right|,
\end{equation}
with the average magnitude of the rung matrix elements ( $r=0$ case in Eq.~\ref{eq:Cr_AB}) of the exact Floquet Hamiltonian,
\begin{equation}
\overline{t'}_{\mathrm{exact}}
=\frac{1}{N}\sum_{i=1}^{N}
\left|(H_F)_{i,N+i}\right|.
\end{equation}
Here, $N$ denotes the number of sites on each leg. As shown in Fig.~\ref{fig:cont2}(b), the two quantities deviate appreciably at low frequencies and approach the bare inter-leg hopping amplitude ($t'$) as the frequency increases. The low-frequency discrepancy indicates the limitations of the effective description in this regime. In particular, the rotating-frame expression predicts a strongly suppressed effective rung coupling when the magnitude of the $\text{sinc}$ factor is small. In this limit, the effective model approaches weakly coupled legs, comprising a clean chain and an Aubry-Andr\'e chain. For $V_B>2t$, the isolated Aubry-Andr\'e chain is localized, whereas the clean chain remains extended. This provides a qualitative interpretation of the coexistence of localized and delocalized states in the effective-model results. The exact Floquet Hamiltonian, however, may not exhibit the same degree of rung-coupling suppression due to the presence of long-range hoppings. In appendix~\ref{long_range_rung}, we explicitly show the presence of long-range interleg hoppings by analyzing the inter-leg block of the Floquet Hamiltonian.

To strengthen our claim, 
we also examine the finite-size scaling of the sorted IPR of the Floquet eigenstates. Figure~\ref{sorted_iprs}(a)-(c) compares three system sizes $2N=600,~1000, ~1500$ at $\omega=14$, $15$, and $17$. At $\omega=14$ and $15$, a finite fraction of the sorted IPR values decreases with increasing system size, while the remaining values exhibit substantially weaker size dependence and remain of order $\mathcal{O}(1)$. This behavior is consistent with the coexistence of extended and localized states for the low-intermediate frequency region. At $\omega=17$, the sorted IPR curves approximately collapse across the system sizes considered, indicating predominantly localized behavior. Similar trends are observed for the effective Hamiltonian in Eq.~\ref{HF0_final}, as shown in Fig.~\ref{sorted_iprs}(d)-(f).

Furthermore, to quantitatively compare the localization transition predicted by the exact Floquet Hamiltonian $H_F$ with those predicted by the high-frequency effective Hamiltonians obtained from the rotating-frame approximation and the van Vleck expansion, we compute the fraction of delocalized states $\bar{n}_c$. 
We define  $\bar n_c$ as the fraction of states that satisfy a finite-size scaling criterion at a given driving frequency. For two system sizes, $2N_1$ and $2N_2$, we sort the IPR values and compare entries with the same normalized rank (so that the state with a low IPR value has a low rank). We define
\begin{equation}
\mathcal{D}_n =
\left|1- \frac{\mathrm{IPR}_{N_1}^{(n_1/2N_1)}}
{\mathrm{IPR}_{N_2}^{(n_2/2N_2)}} \right|,
\qquad
\frac{n_1}{2N_1}=\frac{n_2}{2N_2},
\end{equation}
where $\mathrm{IPR}_{N}^{(n/2N)}$ is the IPR of the state with the normalized rank $n/2N$ with system size $2N$. For $N_2=2N_1$, the matching ranks satisfy $n_2=2n_1$. We use $2N_1=500$ and $2N_2=1000$ in the following analysis.

For extended states, the IPR scales approximately as $1/2N$, so that the ratio of the IPRs at the two system sizes approaches $2$ and $\mathcal{D}_n$ approaches $1$. For localized states, the IPR is approximately independent of system size and order $\mathcal O(1)$, yielding $\mathcal{D}_n\approx0$. We therefore classify states satisfying $\mathcal{D}_n>0.05$ as the presence of mixed phase under our finite-size criterion and define
\begin{equation}
\bar n_c(\omega)=
\frac{1}{2N_1}\sum_{n=1}^{2N_1}
\Theta\left(\mathcal{D}_n(\omega)-0.05\right),
\end{equation}
where $\Theta(x)$ is the Heaviside step function. Thus, $\bar n_c$ measures the fraction of normalized state ranks that satisfy the chosen scaling criterion. The fraction reduces to $\bar n_c = n_c/2N_1$. The threshold $0.05$ is chosen to distinguish size-dependent from approximately size-independent IPR values.

Figure~\ref{fig:n_c_vs_omega} presents $\bar n_c$ as a function of $\omega$ for the exact Floquet Hamiltonian and the effective Hamiltonians. We define $\omega_c$ as the frequency at which $\bar n_c$ vanishes under the above criterion. In the intermediate-frequency regime, the exact and effective descriptions exhibit qualitatively similar trends and yield comparable estimates of the critical frequency. The rotating-frame description agrees more closely with the exact results than the truncated van Vleck description over the parameter range considered. This difference is consistent with retaining the full frequency-dependent $sinc$ renormalization of the inter-leg hopping in the rotating-frame treatment, whereas the van Vleck Hamiltonian is truncated at finite order in the high-frequency expansion. Since the crossover occurs in a regime where the effective descriptions remain reasonably accurate, both approaches capture its approximate frequency scale, despite their differences at lower frequencies.

\section{Conclusion} \label{conc}
In summary, we have investigated the localization properties of a two-leg ladder in which an Aubry-Andr\'e chain is coupled to a clean tight-binding chain subjected to a periodic drive. In contrast to uniform driving of the entire system, we find, both numerically using the exact Floquet Hamiltonian and analytically by examining the high-frequency effective Hamiltonian, that driving only the clean subsystem does not lead to complete delocalization below a critical driving frequency. Instead, a mixed regime emerges in which localized and delocalized states coexist, with the fraction of delocalized states exhibiting a nonmonotonic dependence on the driving frequency. These results demonstrate that selective driving can modify the localization properties of a composite system in a qualitatively different manner from global driving.
\par
Thus, our results establish subsystem-selective driving as an alternative route to Floquet control of localization, in which the transport properties of the full system can be controlled via the dynamics of a single constituent subsystem. More broadly, they show that the interplay between periodic driving and intersubsystem hybridization can provide a mechanism for engineering localization properties without directly driving the entire lattice.

This scheme also offers practical advantages for experimental realization. Since the drive acts only on the clean chain, the quasiperiodic potential need not be modulated, and the drive can be applied through a single, spatially uniform control field on one leg. Both quasiperiodic chains and two-leg ladder geometries have been realized in ultracold atoms, photonic waveguide arrays, and superconducting qubit arrays, where leg-selective modulation can be implemented through local addressing, for example by modulating the propagation constants of one row of waveguides or the frequencies of one row of qubits. Our predictions, in particular the coexistence of localized and delocalized states and the nonmonotonic frequency dependence of the delocalized fraction, could be probed by tracking the spreading of an initially localized wave packet launched on either leg.

Several directions merit further investigation. A natural extension is to include interactions, where the interplay between many-body localization, inter-leg hybridization, and subsystem-selective driving remains an open question. In particular, it would be interesting to examine whether the mixed regime found here persists in the presence of interactions, possibly giving rise to many-body mobility edges, and whether restricting the drive to the clean subsystem can suppress the drive-induced heating that generically destabilizes localization in interacting Floquet systems. Other promising extensions include ladders with more than two legs, quasiperiodic or random driving protocols, and the role of the inter-leg coupling strength beyond the high-frequency regime, where nonlocal effective couplings become important.

\begin{acknowledgments}
R.M. acknowledges the DST-Inspire fellowship from the
Department of Science and Technology, Government of India.
\end{acknowledgments}

\appendix
\section{Localization of the undriven Ladder system} \label{static_loc}
\begin{figure}[h!]
    \centering
    \includegraphics[width=0.45\textwidth]{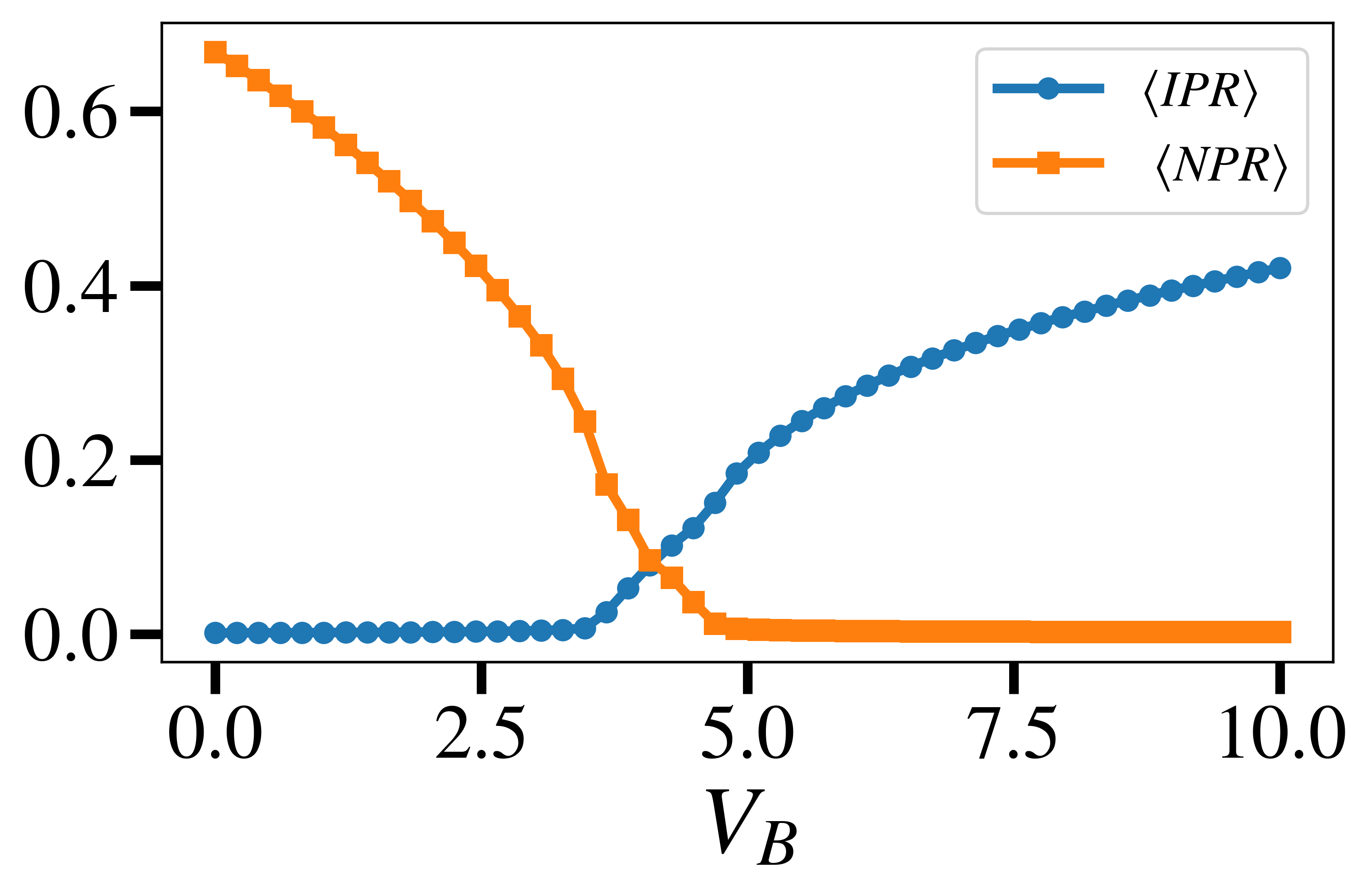}
    \caption{$\langle IPR \rangle$, $\langle NPR \rangle$ as a function of $V_B$ for $t'=10$ and $t_A=t_B =1$ for the undriven Hamiltonian.}
    \label{fig:ipr_npr}
\end{figure}
We first establish the localization properties of the static two-leg ladder by considering the undriven limit, $f_A(t)=0$, of Hamiltonian Eq.~\ref{ham_eqn}. The static localization properties are primarily controlled by the onsite potential strength $V_B$ on leg $B$ and the interleg coupling amplitude $t'$. In the decoupled limit $t'=0$, leg $A$ corresponds to a clean tight-binding chain and therefore supports extended states for all energies. In contrast, leg $B$ is described by the Aubry-Andr\'e model and undergoes a localization transition at $V_B=2t$. Thus, for $V_B<2t$, both legs support extended states, whereas for $V_B>2t$, the states associated with leg $B$ become localized while those associated with the clean leg $A$ remain extended. In the thermodynamic limit, this corresponds to approximately half of the spectrum being localized and the other half remaining extended in the decoupled limit.

A finite interleg coupling hybridizes the two legs, thereby modifying this simple picture. In particular, sufficiently strong interleg coupling can drive the coupled ladder into an overall localized regime through bond localization~\cite{subsystem_localization}. Figure~\ref{fig:ipr_npr} confirms that our choice of parameters, $t'=10$ and $V_B=6$, places the undriven system deep in the localized phase, where the eigenstate-averaged $\langle \mathrm{IPR}\rangle$ remains finite while $\langle \mathrm{NPR}\rangle$ vanishes.

\begin{figure*}[t]
    \centering
    \includegraphics[width=0.96\textwidth]{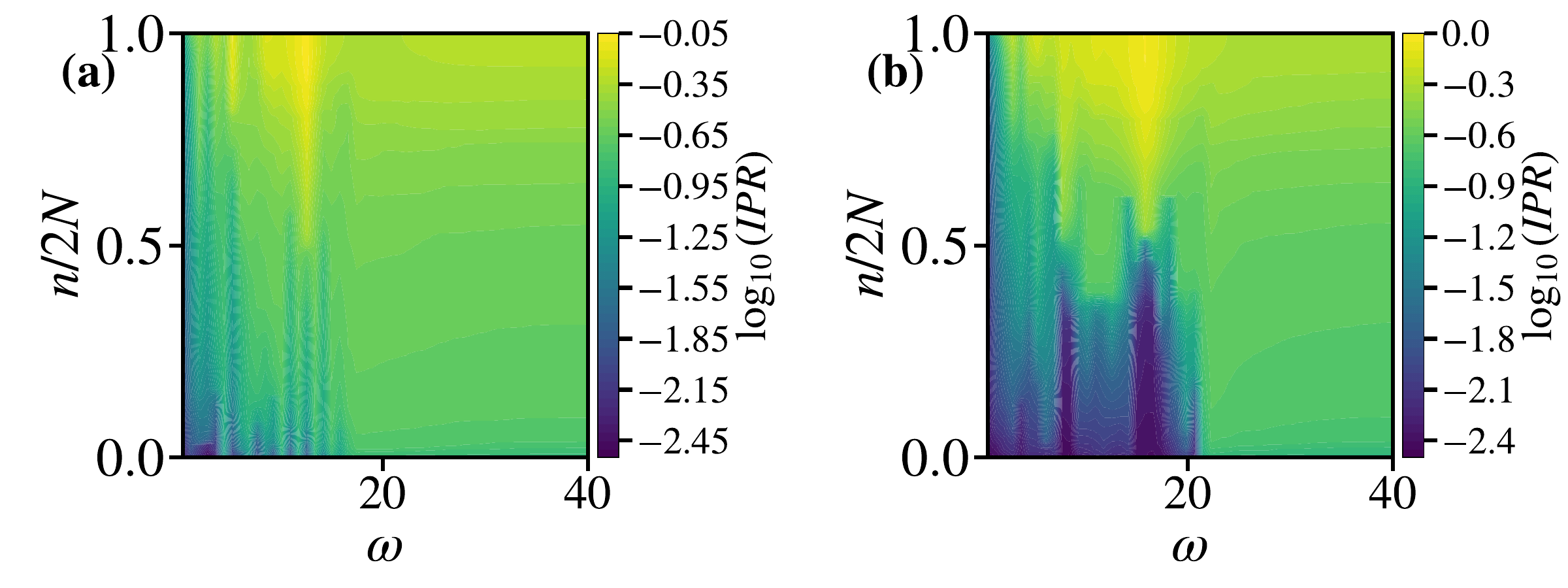}
    \caption{(a) Contour plot of sorted $\log{(IPR)}$ as a function of normalized state index and $\omega$ for (a) $V_A=15$ and (b) $V_B =25$ for $V_B=6$, $t'=10$, and total system size $2N=1000$. }
    \label{fig:cont_app}
\end{figure*}

\section{Contour plots of sorted IPR  for $V_A =15$ and $V_A=25$ } \label{app_3}

To further illustrate the dependence of localization on the strength of the square drive $V_A$, we present the sorted IPR contours for $V_A=15$ and $25$ in Fig.~\ref{fig:cont_app}. A higher $V_A$ indicates a larger fraction of delocalized states in the low-$\omega$ region. Interestingly, the critical frequency $\omega_c$ for the mixed-to-localized transition shifts toward higher frequency, and the range of $\omega$ over which mixed states survive also increases with $V_A$.

\par This behavior can be understood using the analytically derived $t'_{\mathrm{eff}}$ formula in Eq.~\ref{teff}. Since $t'_{\mathrm{eff}} = t' \mathrm{sinc}(V_A T/4)$, we note that $\mathrm{sinc}(V_AT/4) \to 1$ for $V_AT/4 \to 0$, i.e., when $\pi V_A / 2\omega \to 0$. Thus, to reach the static limit ($t'_{\mathrm{eff}} \to t'$), one requires a higher value of $\omega$ for larger $V_A$, ensuring that the ratio $\pi V_A / 2\omega$ becomes small.

\section{Higher-order van Vleck expansion}
\label{app:van_vleck}
To complement the effective-Hamiltonian analysis presented in the main text, we provide here the higher-order van Vleck expansion for the square-wave-driven ladder. The time-dependent Hamiltonian is written as
\begin{equation}
H(t)=\sum_{m=-\infty}^{\infty}H_m e^{im\omega t},
\end{equation}
where
\begin{equation}
H_0=H_A+H_B+H_{AB},
\qquad
H_m=f_mN_A,\quad m\neq0.
\end{equation}
For the square-wave protocol,
\begin{equation}
f_m=
\begin{cases}
\dfrac{2V_A}{i\pi m}, & m~\mathrm{odd},\\[6pt]
0, & m~\mathrm{even}.
\end{cases}
\end{equation}
The effective van Vleck Hamiltonian is expanded as
\begin{equation}
H_{\rm vV} =
H^{(0)}+H^{(1)}+H^{(2)}+H^{(3)} +\mathcal{O}(\omega^{-4}).
\end{equation}
The zeroth-order contribution is simply
\begin{equation}
H^{(0)}=H_0.
\end{equation}
The first-order contribution is
\begin{equation}
H^{(1)} = \sum_{m\neq0} \frac{[H_{-m},H_m]}{2m\omega}.
\end{equation}
Since all the  nonzero Fourier components are proportional to the same
operator $N_A$, one has
\begin{equation}
[H_m,H_n] = f_mf_n[N_A,N_A] = 0.
\end{equation}
Consequently,
\begin{equation}
H^{(1)}=0.
\end{equation}
The leading nonvanishing contribution is therefore the second-order term,
\begin{equation}
\begin{split}
H^{(2)} & =
\sum_{m\neq0}
\frac{[H_{-m},[H_0,H_m]]} 
{2(m\omega)^2} \\& + \sum_{m \ne 0} \sum_{n \ne0,m} \frac{[[H_{-m}, H_{m-n}], H_n]}{3m^2n\omega^2}.
\end{split}
\end{equation}
Using $H_m=f_mN_A$, this becomes
\begin{equation}
H^{(2)} = \sum_{m\neq0}
\frac{|f_m|^2}{2m^2\omega^2} [N_A,[H_0,N_A]],
\end{equation}
The second term goes to 0 as:
\[\frac{[[H_{-m}, H_{m-n}], H_n]}{3m^2n\omega^2} =f_{-m} f_{m-n} f_{n}\frac{[[N_A, N_A], N_A]}{3m^2n\omega^2} \]

For the square-wave drive,
\begin{equation}
|f_m|^2=\frac{4V_A^2}{\pi^2m^2},
\end{equation}
and hence
\begin{equation}
H^{(2)} =\frac{2V_A^2}{\pi^2\omega^2}
\left(\sum_{m~{\rm odd}}\frac{1}{m^4}\right)
[N_A,[H_0,N_A]].
\end{equation}
Using
\begin{equation}
\sum_{m~{\rm \pm 1, \pm3, \dots}}\frac{1}{m^4}
=\frac{\pi^4}{48},
\end{equation}
we obtain
\begin{equation}\label{app:H2final}
H^{(2)} = \frac{\pi^2V_A^2}{24\omega^2}
[N_A,[H_0,N_A]].
\end{equation}
Because $N_A$ commutes with both intraleg Hamiltonians $H_A$ and $H_B$, the nested commutator in Eq.~\eqref{app:H2final} involves only the interleg hopping term $H_{AB}$. Thus, the second-order van Vleck correction preserves the local ladder structure and only renormalizes the rung hopping amplitude. This is consistent with the effective Hamiltonian obtained from the rotating-frame approach in the high-frequency regime.

Now as we have \[H_0 =H_A +H_B +H_{AB}.\] 
Therefore, \[[N_A,[H_0,N_A]]=[N_A,[H_{AB}, N_A]],\] where \[H_{AB} =-t'\sum_{n} \left(a^{\dagger}_nb_n +b^{\dagger}_n a_n\right)\] 
To evaluate the commutator relations, we need to determine the value of the following commutation:
\[[a_i^{\dagger} a_i, a_j^{\dagger}]=a_i^{\dagger}\delta_{ij}\]
 we use $[a_i, a_j^{\dagger}] =\delta_{ij}$ and $[a_i, a_j]=[a_i^{\dagger}, a_j^{\dagger}]=0.$
Similarly, \[[a_i^{\dagger} a_i, a_j]=-\delta_{ij}a_i\]
Using these relations, we get:
\[[ H_{AB}, N_A]= t'\sum_{n} \left(a_n^{\dagger }b_n-b_n^{\dagger }a_n\right)\]

\[[N_A,[H_{AB}, N_A]]= t'\sum_{n} \left(a_n^{\dagger }b_n+b_n^{\dagger }a_n\right)\]

Therefore, we get the following final effective Hamiltonian expression: \begin{equation}
H_{\rm vV}=H_A +H_B -t'\left(1-\frac{\pi^2 V_A^2}{24\omega^2}\right)\sum_{n}\left(a^{\dagger}_nb_n +b^{\dagger}_n a_n\right)
\end{equation}

\begin{figure*}[t]
    \centering
    \includegraphics[width=1\textwidth]{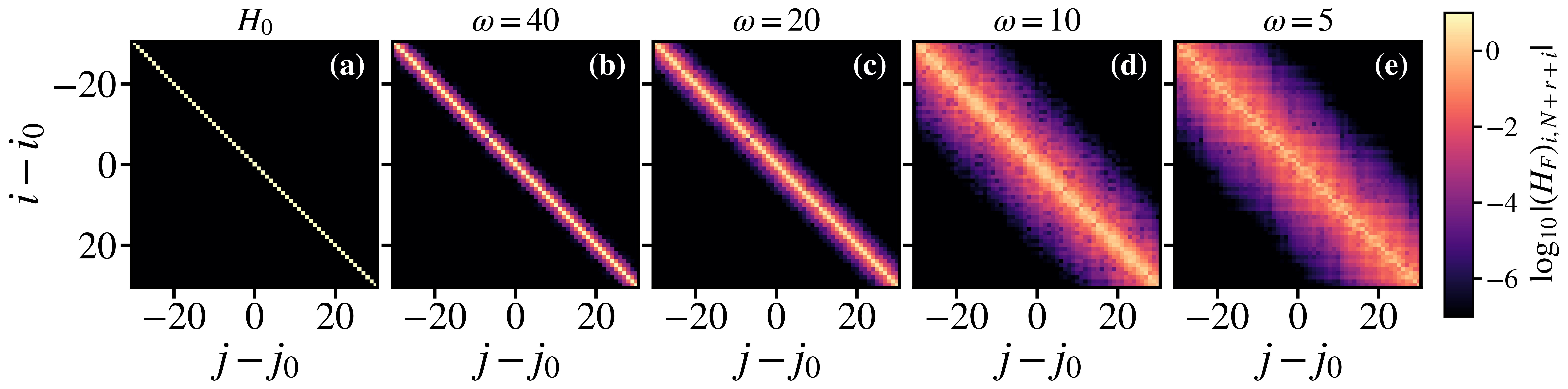}
    \caption{\textcolor{black}{Plot for absolute matrix elements of the inter-leg Hamiltonian block $\log_{10}\left|( H_F)_{i, N+r+i}\right|$ plot for $t'=10$, $V_B=6$, and $V_A=20$ shown in a zoomed in region around $i_0$ and $j_0$ at different driving frequencies: (a) undriven system (or theoretically $\omega \to \infty$), (b) $\omega=40$, (c) $\omega=20$, (d) $\omega=10$, and (e) $\omega=5$. We fixed $2N=2000$ for our analysis.}}
    \label{fig:mat_ele}
\end{figure*}

\section{Matrix structure of the exact Floquet Hamiltonian} \label{long_range_rung}
To see the structural change in the Hamiltonian, we also analyze the logarithmic matrix elements corresponding to the inter-leg hopping block of the Floquet Hamiltonian as $\log_{10}\left|( H_F)_{i, N+r+i}\right|$ (see the discussion in Eq.~\ref{eq:Cr_AB}).
The results can be found in Figure~\ref{fig:mat_ele}. As a starting point, first we plot the inter-leg hopping block of the Hamiltonian matrix corresponding to the static Hamiltonian (see~\ref{fig:mat_ele}(a)). The result shows that the static Hamiltonian had a highly local structure, with dominant matrix elements confined to the main diagonal. At high driving frequency (see Fig.~\ref{fig:mat_ele}(b) and \ref{fig:mat_ele}(c)), the Floquet Hamiltonian exhibits a sharp banded structure, with dominant matrix elements confined to the main diagonal, corresponding to the nearest-neighbor rung hopping terms. As a result, the spatial structure of the exact Floquet Hamiltonian remains essentially mostly identical to that of the static ladder, consistent with the high-frequency effective Hamiltonian derived above.

Now, as the driving frequency decreases ($\omega \sim 10-5$, see the corresponding Fig.~\ref{fig:mat_ele}(d) and Fig.~\ref{fig:mat_ele}(e)), the matrix structure becomes progressively broader. In particular, appreciable matrix elements appear off the diagonals associated with local hopping processes. These additional matrix elements connect sites on different rungs and therefore correspond to effective hopping processes that are absent in the static Hamiltonian and the leading-order Floquet description. The broadening becomes increasingly dominant at intermediate and low frequencies, indicating that the Floquet Hamiltonian acquires an intrinsically nonlocal structure.

\nocite{*}

\bibliography{manuscript}

@article{AL_1,
  title = {Absence of Diffusion in Certain Random Lattices},
  author = {Anderson, P. W.},
  journal = {Phys. Rev.},
  volume = {109},
  issue = {5},
  pages = {1492--1505},
  numpages = {0},
  year = {1958},
  month = {Mar},
  publisher = {American Physical Society},
  doi = {10.1103/PhysRev.109.1492},
  url = {https://link.aps.org/doi/10.1103/PhysRev.109.1492}
}

@article{AL_2,
  title = {Disordered electronic systems},
  author = {Lee, Patrick A. and Ramakrishnan, T. V.},
  journal = {Rev. Mod. Phys.},
  volume = {57},
  issue = {2},
  pages = {287--337},
  numpages = {0},
  year = {1985},
  month = {Apr},
  publisher = {American Physical Society},
  doi = {10.1103/RevModPhys.57.287},
  url = {https://link.aps.org/doi/10.1103/RevModPhys.57.287}
}

@article{SL_1,
  title = {Scaling Theory of Localization: Absence of Quantum Diffusion in Two Dimensions},
  author = {Abrahams, E. and Anderson, P. W. and Licciardello, D. C. and Ramakrishnan, T. V.},
  journal = {Phys. Rev. Lett.},
  volume = {42},
  issue = {10},
  pages = {673--676},
  numpages = {0},
  year = {1979},
  month = {Mar},
  publisher = {American Physical Society},
  doi = {10.1103/PhysRevLett.42.673},
  url = {https://link.aps.org/doi/10.1103/PhysRevLett.42.673}
}

@article{SL_2,
title = {Electrons in disordered systems and the theory of localization},
journal = {Physics Reports},
volume = {13},
number = {3},
pages = {93-142},
year = {1974},
issn = {0370-1573},
doi = {https://doi.org/10.1016/0370-1573(74)90029-5},
url = {https://www.sciencedirect.com/science/article/pii/0370157374900295},
author = {D.J. Thouless}
}

@article{anderson_3d,
  title = {Localization by pseudorandom potentials in one dimension},
  author = {Griniasty, Meir and Fishman, Shmuel},
  journal = {Phys. Rev. Lett.},
  volume = {60},
  issue = {13},
  pages = {1334--1337},
  numpages = {0},
  year = {1988},
  month = {Mar},
  publisher = {American Physical Society},
  doi = {10.1103/PhysRevLett.60.1334},
  url = {https://link.aps.org/doi/10.1103/PhysRevLett.60.1334}
}

@article{QP_1,
  title = {Critical Behavior and Fractality in Shallow One-Dimensional Quasiperiodic Potentials},
  author = {Yao, Hepeng and Khoudli, Alice and Bresque, L\'ea and Sanchez-Palencia, Laurent},
  journal = {Phys. Rev. Lett.},
  volume = {123},
  issue = {7},
  pages = {070405},
  numpages = {6},
  year = {2019},
  month = {Aug},
  publisher = {American Physical Society},
  doi = {10.1103/PhysRevLett.123.070405},
  url = {https://link.aps.org/doi/10.1103/PhysRevLett.123.070405}
}

@article{QP_2,
  title = {Fraction of delocalized eigenstates in the long-range Aubry-Andr\'e-Harper model},
  author = {Roy, Nilanjan and Sharma, Auditya},
  journal = {Phys. Rev. B},
  volume = {103},
  issue = {7},
  pages = {075124},
  numpages = {10},
  year = {2021},
  month = {Feb},
  publisher = {American Physical Society},
  doi = {10.1103/PhysRevB.103.075124},
  url = {https://link.aps.org/doi/10.1103/PhysRevB.103.075124}
}

@article{QP_3,
  title = {One-Dimensional Quasiperiodic Mosaic Lattice with Exact Mobility Edges},
  author = {Wang, Yucheng and Xia, Xu and Zhang, Long and Yao, Hepeng and Chen, Shu and You, Jiangong and Zhou, Qi and Liu, Xiong-Jun},
  journal = {Phys. Rev. Lett.},
  volume = {125},
  issue = {19},
  pages = {196604},
  numpages = {6},
  year = {2020},
  month = {Nov},
  publisher = {American Physical Society},
  doi = {10.1103/PhysRevLett.125.196604},
  url = {https://link.aps.org/doi/10.1103/PhysRevLett.125.196604}
}

@article{QP_4,
  title = {Reentrant Localization Transition in a Quasiperiodic Chain},
  author = {Roy, Shilpi and Mishra, Tapan and Tanatar, B. and Basu, Saurabh},
  journal = {Phys. Rev. Lett.},
  volume = {126},
  issue = {10},
  pages = {106803},
  numpages = {5},
  year = {2021},
  month = {Mar},
  publisher = {American Physical Society},
  doi = {10.1103/PhysRevLett.126.106803},
  url = {https://link.aps.org/doi/10.1103/PhysRevLett.126.106803}
}

@Article{QP_5,
	title={{Anomalous mobility edges in one-dimensional quasiperiodic models}},
	author={Tong Liu and Xu Xia and Stefano Longhi and Laurent Sanchez-Palencia},
	journal={SciPost Phys.},
	volume={12},
	pages={027},
	year={2022},
	publisher={SciPost},
	doi={10.21468/SciPostPhys.12.1.027},
	url={https://scipost.org/10.21468/SciPostPhys.12.1.027},
}

@article{QP_6,
  title = {Localization and multifractal properties of the long-range Kitaev chain in the presence of an Aubry-Andr\'e-Harper modulation},
  author = {Fraxanet, Joana and Bhattacharya, Utso and Grass, Tobias and Lewenstein, Maciej and Dauphin, Alexandre},
  journal = {Phys. Rev. B},
  volume = {106},
  issue = {2},
  pages = {024204},
  numpages = {12},
  year = {2022},
  month = {Jul},
  publisher = {American Physical Society},
  doi = {10.1103/PhysRevB.106.024204},
  url = {https://link.aps.org/doi/10.1103/PhysRevB.106.024204}
}

@Article{QP_7,
	title={{Hidden dualities in 1D quasiperiodic lattice models}},
	author={Miguel Gonçalves and Bruno Amorim and Eduardo V. Castro and Pedro Ribeiro},
	journal={SciPost Phys.},
	volume={13},
	pages={046},
	year={2022},
	publisher={SciPost},
	doi={10.21468/SciPostPhys.13.3.046},
	url={https://scipost.org/10.21468/SciPostPhys.13.3.046},
}

@article{QP_8,
  title = {Realization and Detection of Nonergodic Critical Phases in an Optical Raman Lattice},
  author = {Wang, Yucheng and Zhang, Long and Niu, Sen and Yu, Dapeng and Liu, Xiong-Jun},
  journal = {Phys. Rev. Lett.},
  volume = {125},
  issue = {7},
  pages = {073204},
  numpages = {7},
  year = {2020},
  month = {Aug},
  publisher = {American Physical Society},
  doi = {10.1103/PhysRevLett.125.073204},
  url = {https://link.aps.org/doi/10.1103/PhysRevLett.125.073204}
}

@article{QP_9,
  title = {Quantum phase with coexisting localized, extended, and critical zones},
  author = {Wang, Yucheng and Zhang, Long and Sun, Wei and Poon, Ting-Fung Jeffrey and Liu, Xiong-Jun},
  journal = {Phys. Rev. B},
  volume = {106},
  issue = {14},
  pages = {L140203},
  numpages = {7},
  year = {2022},
  month = {Oct},
  publisher = {American Physical Society},
  doi = {10.1103/PhysRevB.106.L140203},
  url = {https://link.aps.org/doi/10.1103/PhysRevB.106.L140203}
}

@article{QP_10,
  title = {One-Dimensional Quasicrystals with Power-Law Hopping},
  author = {Deng, X. and Ray, S. and Sinha, S. and Shlyapnikov, G. V. and Santos, L.},
  journal = {Phys. Rev. Lett.},
  volume = {123},
  issue = {2},
  pages = {025301},
  numpages = {6},
  year = {2019},
  month = {Jul},
  publisher = {American Physical Society},
  doi = {10.1103/PhysRevLett.123.025301},
  url = {https://link.aps.org/doi/10.1103/PhysRevLett.123.025301}
}

@article{QP_11,
  title = {Mobility edges in one-dimensional bichromatic incommensurate potentials},
  author = {Li, Xiao and Li, Xiaopeng and Das Sarma, S.},
  journal = {Phys. Rev. B},
  volume = {96},
  issue = {8},
  pages = {085119},
  numpages = {18},
  year = {2017},
  month = {Aug},
  publisher = {American Physical Society},
  doi = {10.1103/PhysRevB.96.085119},
  url = {https://link.aps.org/doi/10.1103/PhysRevB.96.085119}
}

@article{EXP_1,
  title = {Interactions and Mobility Edges: Observing the Generalized Aubry-Andr\'e Model},
  author = {An, Fangzhao Alex and Padavi\ifmmode \acute{c}\else \'{c}\fi{}, Karmela and Meier, Eric J. and Hegde, Suraj and Ganeshan, Sriram and Pixley, J. H. and Vishveshwara, Smitha and Gadway, Bryce},
  journal = {Phys. Rev. Lett.},
  volume = {126},
  issue = {4},
  pages = {040603},
  numpages = {6},
  year = {2021},
  month = {Jan},
  publisher = {American Physical Society},
  doi = {10.1103/PhysRevLett.126.040603},
  url = {https://link.aps.org/doi/10.1103/PhysRevLett.126.040603}
}

@article{EXP_2,
  title = {Ultracold Atoms in a Disordered Crystal of Light: Towards a Bose Glass},
  author = {Fallani, L. and Lye, J. E. and Guarrera, V. and Fort, C. and Inguscio, M.},
  journal = {Phys. Rev. Lett.},
  volume = {98},
  issue = {13},
  pages = {130404},
  numpages = {4},
  year = {2007},
  month = {Mar},
  publisher = {American Physical Society},
  doi = {10.1103/PhysRevLett.98.130404},
  url = {https://link.aps.org/doi/10.1103/PhysRevLett.98.130404}
}

@article{EXP_3,
  title = {Observation of a Disordered Bosonic Insulator from Weak to Strong Interactions},
  author = {D'Errico, Chiara and Lucioni, Eleonora and Tanzi, Luca and Gori, Lorenzo and Roux, Guillaume and McCulloch, Ian P. and Giamarchi, Thierry and Inguscio, Massimo and Modugno, Giovanni},
  journal = {Phys. Rev. Lett.},
  volume = {113},
  issue = {9},
  pages = {095301},
  numpages = {5},
  year = {2014},
  month = {Aug},
  publisher = {American Physical Society},
  doi = {10.1103/PhysRevLett.113.095301},
  url = {https://link.aps.org/doi/10.1103/PhysRevLett.113.095301}
}

@article{EXP_4,
author = {Michael Schreiber  and Sean S. Hodgman  and Pranjal Bordia  and Henrik P. Lüschen  and Mark H. Fischer  and Ronen Vosk  and Ehud Altman  and Ulrich Schneider  and Immanuel Bloch },
title = {Observation of many-body localization of interacting fermions in a quasirandom optical lattice},
journal = {Science},
volume = {349},
number = {6250},
pages = {842-845},
year = {2015},
doi = {10.1126/science.aaa7432},
URL = {https://www.science.org/doi/abs/10.1126/science.aaa7432},
eprint = {https://www.science.org/doi/pdf/10.1126/science.aaa7432}}

@article{EXP_5,
  title = {Coupling Identical one-dimensional Many-Body Localized Systems},
  author = {Bordia, Pranjal and L\"uschen, Henrik P. and Hodgman, Sean S. and Schreiber, Michael and Bloch, Immanuel and Schneider, Ulrich},
  journal = {Phys. Rev. Lett.},
  volume = {116},
  issue = {14},
  pages = {140401},
  numpages = {6},
  year = {2016},
  month = {Apr},
  publisher = {American Physical Society},
  doi = {10.1103/PhysRevLett.116.140401},
  url = {https://link.aps.org/doi/10.1103/PhysRevLett.116.140401}
}

@article{EXP_6,
  title = {Probing Slow Relaxation and Many-Body Localization in Two-Dimensional Quasiperiodic Systems},
  author = {Bordia, Pranjal and L\"uschen, Henrik and Scherg, Sebastian and Gopalakrishnan, Sarang and Knap, Michael and Schneider, Ulrich and Bloch, Immanuel},
  journal = {Phys. Rev. X},
  volume = {7},
  issue = {4},
  pages = {041047},
  numpages = {8},
  year = {2017},
  month = {Nov},
  publisher = {American Physical Society},
  doi = {10.1103/PhysRevX.7.041047},
  url = {https://link.aps.org/doi/10.1103/PhysRevX.7.041047}
}

@article{EXP_7,
  title = {Observation of Slow Dynamics near the Many-Body Localization Transition in One-Dimensional Quasiperiodic Systems},
  author = {L\"uschen, Henrik P. and Bordia, Pranjal and Scherg, Sebastian and Alet, Fabien and Altman, Ehud and Schneider, Ulrich and Bloch, Immanuel},
  journal = {Phys. Rev. Lett.},
  volume = {119},
  issue = {26},
  pages = {260401},
  numpages = {6},
  year = {2017},
  month = {Dec},
  publisher = {American Physical Society},
  doi = {10.1103/PhysRevLett.119.260401},
  url = {https://link.aps.org/doi/10.1103/PhysRevLett.119.260401}
}

@article{EXP_8,
  title = {Engineering a Flux-Dependent Mobility Edge in Disordered Zigzag Chains},
  author = {An, Fangzhao Alex and Meier, Eric J. and Gadway, Bryce},
  journal = {Phys. Rev. X},
  volume = {8},
  issue = {3},
  pages = {031045},
  numpages = {11},
  year = {2018},
  month = {Aug},
  publisher = {American Physical Society},
  doi = {10.1103/PhysRevX.8.031045},
  url = {https://link.aps.org/doi/10.1103/PhysRevX.8.031045}
}

@article{EXP_9,
  title = {Observation of Interaction-Induced Mobility Edge in an Atomic Aubry-Andr\'e Wire},
  author = {Wang, Yunfei and Zhang, Jia-Hui and Li, Yuqing and Wu, Jizhou and Liu, Wenliang and Mei, Feng and Hu, Ying and Xiao, Liantuan and Ma, Jie and Chin, Cheng and Jia, Suotang},
  journal = {Phys. Rev. Lett.},
  volume = {129},
  issue = {10},
  pages = {103401},
  numpages = {7},
  year = {2022},
  month = {Sep},
  publisher = {American Physical Society},
  doi = {10.1103/PhysRevLett.129.103401},
  url = {https://link.aps.org/doi/10.1103/PhysRevLett.129.103401}
}

@article{expt_1,
  title = {Observation of a Localization Transition in Quasiperiodic Photonic Lattices},
  author = {Lahini, Y. and Pugatch, R. and Pozzi, F. and Sorel, M. and Morandotti, R. and Davidson, N. and Silberberg, Y.},
  journal = {Phys. Rev. Lett.},
  volume = {103},
  issue = {1},
  pages = {013901},
  numpages = {4},
  year = {2009},
  month = {Jun},
  publisher = {American Physical Society},
  doi = {10.1103/PhysRevLett.103.013901},
  url = {https://link.aps.org/doi/10.1103/PhysRevLett.103.013901}
}

@article{expt_2,
  title = {Quasiperiodic GaAs-AlAs Heterostructures},
  author = {Merlin, R. and Bajema, K. and Clarke, Roy and Juang, F. -Y. and Bhattacharya, P. K.},
  journal = {Phys. Rev. Lett.},
  volume = {55},
  issue = {17},
  pages = {1768--1770},
  numpages = {0},
  year = {1985},
  month = {Oct},
  publisher = {American Physical Society},
  doi = {10.1103/PhysRevLett.55.1768},
  url = {https://link.aps.org/doi/10.1103/PhysRevLett.55.1768}
}

@article{expt_3,
  title = {Fractal Energy Spectrum of a Polariton Gas in a Fibonacci Quasiperiodic Potential},
  author = {Tanese, D. and Gurevich, E. and Baboux, F. and Jacqmin, T. and Lema\^{\i}tre, A. and Galopin, E. and Sagnes, I. and Amo, A. and Bloch, J. and Akkermans, E.},
  journal = {Phys. Rev. Lett.},
  volume = {112},
  issue = {14},
  pages = {146404},
  numpages = {5},
  year = {2014},
  month = {Apr},
  publisher = {American Physical Society},
  doi = {10.1103/PhysRevLett.112.146404},
  url = {https://link.aps.org/doi/10.1103/PhysRevLett.112.146404}
}

@article{expt_4,
author = {P. Roushan  and C. Neill  and J. Tangpanitanon  and V. M. Bastidas  and A. Megrant  and R. Barends  and Y. Chen  and Z. Chen  and B. Chiaro  and A. Dunsworth  and A. Fowler  and B. Foxen  and M. Giustina  and E. Jeffrey  and J. Kelly  and E. Lucero  and J. Mutus  and M. Neeley  and C. Quintana  and D. Sank  and A. Vainsencher  and J. Wenner  and T. White  and H. Neven  and D. G. Angelakis  and J. Martinis },
title = {Spectroscopic signatures of localization with interacting photons in superconducting qubits},
journal = {Science},
volume = {358},
number = {6367},
pages = {1175-1179},
year = {2017},
doi = {10.1126/science.aao1401},
URL = {https://www.science.org/doi/abs/10.1126/science.aao1401},
eprint = {https://www.science.org/doi/pdf/10.1126/science.aao1401}}

@article{sc_qubit_1,
  author    = {Goblot, V. and {\v{S}}trkalj, A. and Pernet, N. and Lado, J. L. and Dorow, C. and Lema{\^i}tre, A. and Le Gratiet, L. and Harouri, A. and Sagnes, I. and Ravets, S. and Amo, A. and Bloch, J. and Zilberberg, O.},
  title     = {Emergence of criticality through a cascade of delocalization transitions in quasiperiodic chains},
  journal   = {Nature Physics},
  year      = {2020},
  volume    = {16},
  number    = {8},
  pages     = {832--836},
  doi       = {10.1038/s41567-020-0908-7},
  url       = {https://doi.org/10.1038/s41567-020-0908-7},
  publisher = {Nature Publishing Group},
  issn      = {1745-2481}
}

@article{sc_qubit_2,
  author    = {Li, Hao and Wang, Yong-Yi and Shi, Yun-Hao and Huang, Kaixuan and Song, Xiaohui and Liang, Gui-Han and Mei, Zheng-Yang and Zhou, Bozhen and Zhang, He and Zhang, Jia-Chi and Chen, Shu and Zhao, S. P. and Tian, Ye and Yang, Zhan-Ying and Xiang, Zhongcheng and Xu, Kai and Zheng, Dongning and Fan, Heng},
  title     = {Observation of critical phase transition in a generalized Aubry-Andr{\'e}-Harper model with superconducting circuits},
  journal   = {npj Quantum Information},
  year      = {2023},
  volume    = {9},
  number    = {1},
  pages     = {40},
  doi       = {10.1038/s41534-023-00712-w},
  url       = {https://doi.org/10.1038/s41534-023-00712-w},
  publisher = {Nature Publishing Group},
  issn      = {2056-6387}
}

@article{critical_1,
  title = {Energy spectrum and the quantum Hall effect on the square lattice with next-nearest-neighbor hopping},
  author = {Hatsugai, Y. and Kohmoto, M.},
  journal = {Phys. Rev. B},
  volume = {42},
  issue = {13},
  pages = {8282--8294},
  numpages = {0},
  year = {1990},
  month = {Nov},
  publisher = {American Physical Society},
  doi = {10.1103/PhysRevB.42.8282},
  url = {https://link.aps.org/doi/10.1103/PhysRevB.42.8282}
}

@article{critical_2,
  title = {Critical and bicritical properties of Harper's equation with next-nearest-neighbor coupling},
  author = {Han, J. H. and Thouless, D. J. and Hiramoto, H. and Kohmoto, M.},
  journal = {Phys. Rev. B},
  volume = {50},
  issue = {16},
  pages = {11365--11380},
  numpages = {0},
  year = {1994},
  month = {Oct},
  publisher = {American Physical Society},
  doi = {10.1103/PhysRevB.50.11365},
  url = {https://link.aps.org/doi/10.1103/PhysRevB.50.11365}
}

@article{spec_stat_1,
  title = {Statistics of spectra for critical quantum chaos in one-dimensional quasiperiodic systems},
  author = {Takada, Yoshihiro and Ino, Kazusumi and Yamanaka, Masanori},
  journal = {Phys. Rev. E},
  volume = {70},
  issue = {6},
  pages = {066203},
  numpages = {6},
  year = {2004},
  month = {Dec},
  publisher = {American Physical Society},
  doi = {10.1103/PhysRevE.70.066203},
  url = {https://link.aps.org/doi/10.1103/PhysRevE.70.066203}
}

@article{critical_3,
  title = {Localization and adiabatic pumping in a generalized Aubry-Andr\'e-Harper model},
  author = {Liu, Fangli and Ghosh, Somnath and Chong, Y. D.},
  journal = {Phys. Rev. B},
  volume = {91},
  issue = {1},
  pages = {014108},
  numpages = {9},
  year = {2015},
  month = {Jan},
  publisher = {American Physical Society},
  doi = {10.1103/PhysRevB.91.014108},
  url = {https://link.aps.org/doi/10.1103/PhysRevB.91.014108}
}

@article{critical_4,
  title = {Phase diagram of a non-Abelian Aubry-Andr\'e-Harper model with $p$-wave superfluidity},
  author = {Wang, Jun and Liu, Xia-Ji and Xianlong, Gao and Hu, Hui},
  journal = {Phys. Rev. B},
  volume = {93},
  issue = {10},
  pages = {104504},
  numpages = {7},
  year = {2016},
  month = {Mar},
  publisher = {American Physical Society},
  doi = {10.1103/PhysRevB.93.104504},
  url = {https://link.aps.org/doi/10.1103/PhysRevB.93.104504}
}

@article{spec_stat_3,
  title = {New class of level statistics in quantum systems with unbounded diffusion},
  author = {Geisel, T. and Ketzmerick, R. and Petschel, G.},
  journal = {Phys. Rev. Lett.},
  volume = {66},
  issue = {13},
  pages = {1651--1654},
  numpages = {0},
  year = {1991},
  month = {Apr},
  publisher = {American Physical Society},
  doi = {10.1103/PhysRevLett.66.1651},
  url = {https://link.aps.org/doi/10.1103/PhysRevLett.66.1651}
}

@article{spec_stat_4,
  title = {Quantum energy spectra and one-dimensional quasiperiodic systems},
  author = {Machida, Kazushige and Fujita, Mitsutaka},
  journal = {Phys. Rev. B},
  volume = {34},
  issue = {10},
  pages = {7367--7370},
  numpages = {0},
  year = {1986},
  month = {Nov},
  publisher = {American Physical Society},
  doi = {10.1103/PhysRevB.34.7367},
  url = {https://link.aps.org/doi/10.1103/PhysRevB.34.7367}
}

@article{spec_stat_5,
  title = {Anomalous Thouless energy and critical statistics on the metallic side of the many-body localization transition},
  author = {Bertrand, Corentin L. and Garc\'{\i}a-Garc\'{\i}a, Antonio M.},
  journal = {Phys. Rev. B},
  volume = {94},
  issue = {14},
  pages = {144201},
  numpages = {12},
  year = {2016},
  month = {Oct},
  publisher = {American Physical Society},
  doi = {10.1103/PhysRevB.94.144201},
  url = {https://link.aps.org/doi/10.1103/PhysRevB.94.144201}
}

@article{frac_1,
  title = {Fractal measures and their singularities: The characterization of strange sets},
  author = {Halsey, Thomas C. and Jensen, Mogens H. and Kadanoff, Leo P. and Procaccia, Itamar and Shraiman, Boris I.},
  journal = {Phys. Rev. A},
  volume = {33},
  issue = {2},
  pages = {1141--1151},
  numpages = {0},
  year = {1986},
  month = {Feb},
  publisher = {American Physical Society},
  doi = {10.1103/PhysRevA.33.1141},
  url = {https://link.aps.org/doi/10.1103/PhysRevA.33.1141}
}

@article{frac_2,
  title = {Exact Relations between Multifractal Exponents at the Anderson Transition},
  author = {Mirlin, A. D. and Fyodorov, Y. V. and Mildenberger, A. and Evers, F.},
  journal = {Phys. Rev. Lett.},
  volume = {97},
  issue = {4},
  pages = {046803},
  numpages = {4},
  year = {2006},
  month = {Jul},
  publisher = {American Physical Society},
  doi = {10.1103/PhysRevLett.97.046803},
  url = {https://link.aps.org/doi/10.1103/PhysRevLett.97.046803}
}

@article{frac_3,
  title = {Two Scenarios for Quantum Multifractality Breakdown},
  author = {Dubertrand, R. and Garc\'{\i}a-Mata, I. and Georgeot, B. and Giraud, O. and Lemari\'e, G. and Martin, J.},
  journal = {Phys. Rev. Lett.},
  volume = {112},
  issue = {23},
  pages = {234101},
  numpages = {5},
  year = {2014},
  month = {Jun},
  publisher = {American Physical Society},
  doi = {10.1103/PhysRevLett.112.234101},
  url = {https://link.aps.org/doi/10.1103/PhysRevLett.112.234101}
}

@article{dyn_1,
  title = {What Determines the Spreading of a Wave Packet?},
  author = {Ketzmerick, R. and Kruse, K. and Kraut, S. and Geisel, T.},
  journal = {Phys. Rev. Lett.},
  volume = {79},
  issue = {11},
  pages = {1959--1963},
  numpages = {0},
  year = {1997},
  month = {Sep},
  publisher = {American Physical Society},
  doi = {10.1103/PhysRevLett.79.1959},
  url = {https://link.aps.org/doi/10.1103/PhysRevLett.79.1959}
}

@article{dyn_2,
  title = {Effects of interaction on the diffusion of atomic matter waves in one-dimensional quasiperiodic potentials},
  author = {Larcher, M. and Dalfovo, F. and Modugno, M.},
  journal = {Phys. Rev. A},
  volume = {80},
  issue = {5},
  pages = {053606},
  numpages = {8},
  year = {2009},
  month = {Nov},
  publisher = {American Physical Society},
  doi = {10.1103/PhysRevA.80.053606},
  url = {https://link.aps.org/doi/10.1103/PhysRevA.80.053606}
}

@article{ME_1,
  title = {Proposed Experimental Realization of Anderson Localization in Random and Incommensurate Artificially Layered Systems},
  author = {Das Sarma, S. and Kobayashi, Akiko and Prange, R. E.},
  journal = {Phys. Rev. Lett.},
  volume = {56},
  issue = {12},
  pages = {1280--1283},
  numpages = {0},
  year = {1986},
  month = {Mar},
  publisher = {American Physical Society},
  doi = {10.1103/PhysRevLett.56.1280},
  url = {https://link.aps.org/doi/10.1103/PhysRevLett.56.1280}
}

@article{ME_2,
  title = {Predicted Mobility Edges in One-Dimensional Incommensurate Optical Lattices: An Exactly Solvable Model of Anderson Localization},
  author = {Biddle, J. and Das Sarma, S.},
  journal = {Phys. Rev. Lett.},
  volume = {104},
  issue = {7},
  pages = {070601},
  numpages = {4},
  year = {2010},
  month = {Feb},
  publisher = {American Physical Society},
  doi = {10.1103/PhysRevLett.104.070601},
  url = {https://link.aps.org/doi/10.1103/PhysRevLett.104.070601}
}

@article{ME_3,
  title = {Nearest Neighbor Tight Binding Models with an Exact Mobility Edge in One Dimension},
  author = {Ganeshan, Sriram and Pixley, J. H. and Das Sarma, S.},
  journal = {Phys. Rev. Lett.},
  volume = {114},
  issue = {14},
  pages = {146601},
  numpages = {5},
  year = {2015},
  month = {Apr},
  publisher = {American Physical Society},
  doi = {10.1103/PhysRevLett.114.146601},
  url = {https://link.aps.org/doi/10.1103/PhysRevLett.114.146601}
}

@article{ME_4,
doi = {10.1209/0295-5075/110/57003},
url = {https://doi.org/10.1209/0295-5075/110/57003},
year = {2015},
month = {jun},
publisher = {EDP Sciences, IOP Publishing and Società Italiana di Fisica},
volume = {110},
number = {5},
pages = {57003},
author = {Sun, M. L. and Wang, G. and Li, N. B. and Nakayama, T.},
title = {Localization-delocalization transition in self-dual quasi-periodic lattices},
journal = {Europhysics Letters}
}

@article{ME_5,
  title = {Self-dual quasiperiodic systems with power-law hopping},
  author = {Gopalakrishnan, Sarang},
  journal = {Phys. Rev. B},
  volume = {96},
  issue = {5},
  pages = {054202},
  numpages = {11},
  year = {2017},
  month = {Aug},
  publisher = {American Physical Society},
  doi = {10.1103/PhysRevB.96.054202},
  url = {https://link.aps.org/doi/10.1103/PhysRevB.96.054202}
}

@article{ME_6,
  title = {Nonequilibrium phase diagram of a one-dimensional quasiperiodic system with a single-particle mobility edge},
  author = {Purkayastha, Archak and Dhar, Abhishek and Kulkarni, Manas},
  journal = {Phys. Rev. B},
  volume = {96},
  issue = {18},
  pages = {180204},
  numpages = {5},
  year = {2017},
  month = {Nov},
  publisher = {American Physical Society},
  doi = {10.1103/PhysRevB.96.180204},
  url = {https://link.aps.org/doi/10.1103/PhysRevB.96.180204}
}

@article{ipr_1,
  title = {Electronic Properties of the 1D Frenkel-Kontorova Model},
  author = {Tong, Peiqing and Li, Baowen and Hu, Bambi},
  journal = {Phys. Rev. Lett.},
  volume = {88},
  issue = {4},
  pages = {046804},
  numpages = {4},
  year = {2002},
  month = {Jan},
  publisher = {American Physical Society},
  doi = {10.1103/PhysRevLett.88.046804},
  url = {https://link.aps.org/doi/10.1103/PhysRevLett.88.046804}
}

@article{ipr_2,
title = {Anomalous localization in low-dimensional systems with correlated disorder},
journal = {Physics Reports},
volume = {512},
number = {3},
pages = {125-254},
year = {2012},
note = {Anomalous localization in low-dimensional systems with correlated disorder},
issn = {0370-1573},
doi = {https://doi.org/10.1016/j.physrep.2011.11.002},
url = {https://www.sciencedirect.com/science/article/pii/S0370157311002936},
author = {F.M. Izrailev and A.A. Krokhin and N.M. Makarov}
}

@article{ipr_3,
  title = {Comment on ``Periodic wave functions and number of extended states in random dimer systems''},
  author = {Hilke, M. and Flores, J. C. and Dom\'{\i}nguez-Adame, F.},
  journal = {Phys. Rev. B},
  volume = {58},
  issue = {13},
  pages = {8837--8838},
  numpages = {0},
  year = {1998},
  month = {Oct},
  publisher = {American Physical Society},
  doi = {10.1103/PhysRevB.58.8837},
  url = {https://link.aps.org/doi/10.1103/PhysRevB.58.8837}
}

@article{pr_sc_1,
  title = {Critical analysis of the reentrant localization transition in a one-dimensional dimerized quasiperiodic lattice},
  author = {Roy, Shilpi and Chattopadhyay, Sourav and Mishra, Tapan and Basu, Saurabh},
  journal = {Phys. Rev. B},
  volume = {105},
  issue = {21},
  pages = {214203},
  numpages = {8},
  year = {2022},
  month = {Jun},
  publisher = {American Physical Society},
  doi = {10.1103/PhysRevB.105.214203},
  url = {https://link.aps.org/doi/10.1103/PhysRevB.105.214203}
}

@article{pr_sc_2,
  title = {Anderson transitions},
  author = {Evers, Ferdinand and Mirlin, Alexander D.},
  journal = {Rev. Mod. Phys.},
  volume = {80},
  issue = {4},
  pages = {1355--1417},
  numpages = {0},
  year = {2008},
  month = {Oct},
  publisher = {American Physical Society},
  doi = {10.1103/RevModPhys.80.1355},
  url = {https://link.aps.org/doi/10.1103/RevModPhys.80.1355}
}

@article{nandkishore2015many,
  title={Many-body localization and thermalization in quantum statistical mechanics},
  author={Nandkishore, Rahul and Huse, David A},
  journal={Annu. Rev. Condens. Matter Phys.},
  volume={6},
  number={1},
  pages={15--38},
  year={2015},
  publisher={Annual Reviews}
}

@article{ETH1,
  title = {Quantum statistical mechanics in a closed system},
  author = {Deutsch, J. M.},
  journal = {Phys. Rev. A},
  volume = {43},
  issue = {4},
  pages = {2046--2049},
  numpages = {0},
  year = {1991},
  month = {Feb},
  publisher = {American Physical Society},
  doi = {10.1103/PhysRevA.43.2046},
  url = {https://link.aps.org/doi/10.1103/PhysRevA.43.2046}
}

@article{ETH2,
  title = {Chaos and quantum thermalization},
  author = {Srednicki, Mark},
  journal = {Phys. Rev. E},
  volume = {50},
  issue = {2},
  pages = {888--901},
  numpages = {0},
  year = {1994},
  month = {Aug},
  publisher = {American Physical Society},
  doi = {10.1103/PhysRevE.50.888},
  url = {https://link.aps.org/doi/10.1103/PhysRevE.50.888}
}

@article{rigol2012alternatives,
  title={Alternatives to eigenstate thermalization},
  author={Rigol, Marcos and Srednicki, Mark},
  journal={Physical review letters},
  volume={108},
  number={11},
  pages={110601},
  year={2012},
  publisher={APS}
}

@article{MBL_breakdown1,
  title = {Avalanches and many-body resonances in many-body localized systems},
  author = {Morningstar, Alan and Colmenarez, Luis and Khemani, Vedika and Luitz, David J. and Huse, David A.},
  journal = {Phys. Rev. B},
  volume = {105},
  issue = {17},
  pages = {174205},
  numpages = {20},
  year = {2022},
  month = {May},
  publisher = {American Physical Society},
  doi = {10.1103/PhysRevB.105.174205},
  url = {https://link.aps.org/doi/10.1103/PhysRevB.105.174205}
}

@article{MBL_breakdown2,
  title = {Bath-induced delocalization in interacting disordered spin chains},
  author = {Sels, Dries},
  journal = {Phys. Rev. B},
  volume = {106},
  issue = {2},
  pages = {L020202},
  numpages = {5},
  year = {2022},
  month = {Jul},
  publisher = {American Physical Society},
  doi = {10.1103/PhysRevB.106.L020202},
  url = {https://link.aps.org/doi/10.1103/PhysRevB.106.L020202}
}

@article{vidmar.2020,
  title = {Quantum chaos challenges many-body localization},
  author = {\ifmmode \check{S}\else \v{S}\fi{}untajs, Jan and Bon\ifmmode \check{c}\else \v{c}\fi{}a, Janez and Prosen, Toma\ifmmode \check{z}\else \v{z}\fi{} and Vidmar, Lev},
  journal = {Phys. Rev. E},
  volume = {102},
  issue = {6},
  pages = {062144},
  numpages = {12},
  year = {2020},
  month = {Dec},
  publisher = {American Physical Society},
  doi = {10.1103/PhysRevE.102.062144},
  url = {https://link.aps.org/doi/10.1103/PhysRevE.102.062144}
}

@article{MBL1,
  title = {Many-body localization in the Heisenberg $XXZ$ magnet in a random field},
  author = {\ifmmode \check{Z}\else \v{Z}\fi{}nidari\ifmmode \check{c}\else \v{c}\fi{}, Marko and Prosen, Toma\ifmmode \check{z}\else \v{z}\fi{} and Prelov\ifmmode \check{s}\else \v{s}\fi{}ek, Peter},
  journal = {Phys. Rev. B},
  volume = {77},
  issue = {6},
  pages = {064426},
  numpages = {5},
  year = {2008},
  month = {Feb},
  publisher = {American Physical Society},
  doi = {10.1103/PhysRevB.77.064426},
  url = {https://link.aps.org/doi/10.1103/PhysRevB.77.064426}
}

@article{MBL2,
  title = {Many-body localization phase transition},
  author = {Pal, Arijeet and Huse, David A.},
  journal = {Phys. Rev. B},
  volume = {82},
  issue = {17},
  pages = {174411},
  numpages = {7},
  year = {2010},
  month = {Nov},
  publisher = {American Physical Society},
  doi = {10.1103/PhysRevB.82.174411},
  url = {https://link.aps.org/doi/10.1103/PhysRevB.82.174411}
}

@article{MBL3,
  title = {Unbounded Growth of Entanglement in Models of Many-Body Localization},
  author = {Bardarson, Jens H. and Pollmann, Frank and Moore, Joel E.},
  journal = {Phys. Rev. Lett.},
  volume = {109},
  issue = {1},
  pages = {017202},
  numpages = {5},
  year = {2012},
  month = {Jul},
  publisher = {American Physical Society},
  doi = {10.1103/PhysRevLett.109.017202},
  url = {https://link.aps.org/doi/10.1103/PhysRevLett.109.017202}
}

@article{MBL4,
  title = {Many-body localization edge in the random-field Heisenberg chain},
  author = {Luitz, David J. and Laflorencie, Nicolas and Alet, Fabien},
  journal = {Phys. Rev. B},
  volume = {91},
  issue = {8},
  pages = {081103},
  numpages = {5},
  year = {2015},
  month = {Feb},
  publisher = {American Physical Society},
  doi = {10.1103/PhysRevB.91.081103},
  url = {https://link.aps.org/doi/10.1103/PhysRevB.91.081103}
}

@article{basko2006metal,
  title={Metal--insulator transition in a weakly interacting many-electron system with localized single-particle states},
  author={Basko, Denis M and Aleiner, Igor L and Altshuler, Boris L},
  journal={Annals of physics},
  volume={321},
  number={5},
  pages={1126--1205},
  year={2006},
  publisher={Elsevier}
}

@article{MBL_rev1,
  title = {Colloquium: Many-body localization, thermalization, and entanglement},
  author = {Abanin, Dmitry A. and Altman, Ehud and Bloch, Immanuel and Serbyn, Maksym},
  journal = {Rev. Mod. Phys.},
  volume = {91},
  issue = {2},
  pages = {021001},
  numpages = {26},
  year = {2019},
  month = {May},
  publisher = {American Physical Society},
  doi = {10.1103/RevModPhys.91.021001},
  url = {https://link.aps.org/doi/10.1103/RevModPhys.91.021001}
}

@article{sierant25,
doi = {10.1088/1361-6633/ad9756},
year = {2025},
month = {jan},
publisher = {IOP Publishing},
volume = {88},
number = {2},
pages = {026502},
author = {Sierant, Piotr and Lewenstein, Maciej and Scardicchio, Antonello and Vidmar, Lev and Zakrzewski, Jakub},
title = {Many-body localization in the age of classical computing},
journal = {Reports on Progress in Physics}
}

@article{subsystem_localization,
  title = {Subsystem localization in a two-leg ladder system},
  author = {Goswami, Arpita and Chatterjee, Pallabi and Modak, Ranjan and Sahoo, Shaon},
  journal = {Phys. Rev. B},
  volume = {112},
  issue = {14},
  pages = {144205},
  numpages = {17},
  year = {2025},
  month = {Oct},
  publisher = {American Physical Society},
  doi = {10.1103/95t3-6qd4}
}

@article{Sen_2021,
doi = {10.1088/1361-648X/ac1b61},
url = {https://doi.org/10.1088/1361-648X/ac1b61},
year = {2021},
month = {aug},
publisher = {IOP Publishing},
volume = {33},
number = {44},
pages = {443003},
author = {Sen, Arnab and Sen, Diptiman and Sengupta, K},
title = {Analytic approaches to periodically driven closed quantum systems: methods and applications},
journal = {Journal of Physics: Condensed Matter}
}

@article{Eckardt_2015,
doi = {10.1088/1367-2630/17/9/093039},
url = {https://doi.org/10.1088/1367-2630/17/9/093039},
year = {2015},
month = {sep},
publisher = {IOP Publishing},
volume = {17},
number = {9},
pages = {093039},
author = {Eckardt, André and Anisimovas, Egidijus},
title = {High-frequency approximation for periodically driven quantum systems from a Floquet-space perspective},
journal = {New Journal of Physics}
}

@article{eff_floquet_ham1,
  title = {Periodically Driven Quantum Systems: Effective Hamiltonians and Engineered Gauge Fields},
  author = {Goldman, N. and Dalibard, J.},
  journal = {Phys. Rev. X},
  volume = {4},
  issue = {3},
  pages = {031027},
  numpages = {29},
  year = {2014},
  month = {Aug},
  publisher = {American Physical Society},
  doi = {10.1103/PhysRevX.4.031027},
  url = {https://link.aps.org/doi/10.1103/PhysRevX.4.031027}
}

@article{electric_circuit,
  title = {Observation of non-Abelian Anderson localization and transition in topolectrical circuits},
  author = {Wang, Haiteng and Zhang, Weixuan and Sun, Houjun and Zhang, Xiangdong},
  journal = {Phys. Rev. B},
  volume = {108},
  issue = {14},
  pages = {144203},
  numpages = {14},
  year = {2023},
  month = {Oct},
  publisher = {American Physical Society},
  doi = {10.1103/PhysRevB.108.144203},
  url = {https://link.aps.org/doi/10.1103/PhysRevB.108.144203}
}

@article{peierls,
  title = {Effects of Peierls phases in open linear chains},
  author = {Marques, A. M.},
  journal = {Phys. Rev. B},
  volume = {110},
  issue = {11},
  pages = {115156},
  numpages = {13},
  year = {2024},
  month = {Sep},
  publisher = {American Physical Society},
  doi = {10.1103/PhysRevB.110.115156},
  url = {https://link.aps.org/doi/10.1103/PhysRevB.110.115156}
}

@article{nonergodic_1,
  title = {Many-Body Critical Phase: Extended and Nonthermal},
  author = {Wang, Yucheng and Cheng, Chen and Liu, Xiong-Jun and Yu, Dapeng},
  journal = {Phys. Rev. Lett.},
  volume = {126},
  issue = {8},
  pages = {080602},
  numpages = {6},
  year = {2021},
  month = {Feb},
  publisher = {American Physical Society},
  doi = {10.1103/PhysRevLett.126.080602},
  url = {https://link.aps.org/doi/10.1103/PhysRevLett.126.080602}
}

@article{dyn_loc_1,
  title = {Exploring dynamic localization with a Bose-Einstein condensate},
  author = {Eckardt, Andr\'e and Holthaus, Martin and Lignier, Hans and Zenesini, Alessandro and Ciampini, Donatella and Morsch, Oliver and Arimondo, Ennio},
  journal = {Phys. Rev. A},
  volume = {79},
  issue = {1},
  pages = {013611},
  numpages = {7},
  year = {2009},
  month = {Jan},
  publisher = {American Physical Society},
  doi = {10.1103/PhysRevA.79.013611},
  url = {https://link.aps.org/doi/10.1103/PhysRevA.79.013611}
}

@article{dyn_loc_2,
title = {Dynamic localization of a particle in an electric field viewed in momentum space: Connection with Bloch oscillations},
author = {D.H. Dunlap and V.M. Kenkre},
journal = {Physics Letters A},
volume = {127},
number = {8},
pages = {438-440},
year = {1988},
issn = {0375-9601},
doi = {https://doi.org/10.1016/0375-9601(88)90213-7},
url = {https://www.sciencedirect.com/science/article/pii/0375960188902137}

}

@article{dyn_loc_3,
  title = {Driving induced many-body localization},
  author = {Bairey, Eyal and Refael, Gil and Lindner, Netanel H.},
  journal = {Phys. Rev. B},
  volume = {96},
  issue = {2},
  pages = {020201(R)},
  numpages = {5},
  year = {2017},
  month = {Jul},
  publisher = {American Physical Society},
  doi = {10.1103/PhysRevB.96.020201},
  url = {https://link.aps.org/doi/10.1103/PhysRevB.96.020201}
}

@article{tunn_1,
  title = {Coherent destruction of tunneling},
  author = {Grossmann, F. and Dittrich, T. and Jung, P. and H\"anggi, P.},
  journal = {Phys. Rev. Lett.},
  volume = {67},
  issue = {4},
  pages = {516--519},
  numpages = {0},
  year = {1991},
  month = {Jul},
  publisher = {American Physical Society},
  doi = {10.1103/PhysRevLett.67.516},
  url = {https://link.aps.org/doi/10.1103/PhysRevLett.67.516}
}

@article{tunn_2,
  title = {Coherent destruction of tunneling, dynamic localization, and the Landau-Zener formula},
  author = {Kayanuma, Yosuke and Saito, Keiji},
  journal = {Phys. Rev. A},
  volume = {77},
  issue = {1},
  pages = {010101(R)},
  numpages = {4},
  year = {2008},
  month = {Jan},
  publisher = {American Physical Society},
  doi = {10.1103/PhysRevA.77.010101},
  url = {https://link.aps.org/doi/10.1103/PhysRevA.77.010101}
}

@article{tunn_3,
doi = {10.1088/1367-2630/16/1/013007},
url = {https://doi.org/10.1088/1367-2630/16/1/013007},
year = {2014},
month = {jan},
publisher = {IOP Publishing},
volume = {16},
number = {1},
pages = {013007},
author = {Luo, Xiaobing and Li, Liping and You, Li and Wu, Biao},
title = {Coherent destruction of tunneling and dark Floquet state},
journal = {New Journal of Physics}
}

@article{Floquet_engineered_tp_1,
  title = {Floquet engineering of topological localization transitions and mobility edges in one-dimensional non-Hermitian quasicrystals},
  author = {Zhou, Longwen},
  journal = {Phys. Rev. Res.},
  volume = {3},
  issue = {3},
  pages = {033184},
  numpages = {15},
  year = {2021},
  month = {Aug},
  publisher = {American Physical Society},
  doi = {10.1103/PhysRevResearch.3.033184},
  url = {https://link.aps.org/doi/10.1103/PhysRevResearch.3.033184}
}

@article{Floquet_engineered_tp_2,
  title = {Transport properties of nonequilibrium systems under the application of light: Photoinduced quantum Hall insulators without Landau levels},
  author = {Kitagawa, Takuya and Oka, Takashi and Brataas, Arne and Fu, Liang and Demler, Eugene},
  journal = {Phys. Rev. B},
  volume = {84},
  issue = {23},
  pages = {235108},
  numpages = {13},
  year = {2011},
  month = {Dec},
  publisher = {American Physical Society},
  doi = {10.1103/PhysRevB.84.235108},
  url = {https://link.aps.org/doi/10.1103/PhysRevB.84.235108}
}

@article{Floquet_engineered_tp_3,
  author    = {Lindner, Netanel H. and Refael, Gil and Galitski, Victor},
  title     = {Floquet topological insulator in semiconductor quantum wells},
  journal   = {Nature Physics},
  year      = {2011},
  volume    = {7},
  number    = {6},
  pages     = {490--495},
  month     = {jun},
  doi       = {10.1038/nphys1926},
  url       = {https://doi.org/10.1038/nphys1926}
}

@article{Floquet_engineered_tp_4,
  title = {Floquet topological phases of non-Hermitian systems},
  author = {Wu, Hong and An, Jun-Hong},
  journal = {Phys. Rev. B},
  volume = {102},
  issue = {4},
  pages = {041119(R)},
  numpages = {7},
  year = {2020},
  month = {Jul},
  publisher = {American Physical Society},
  doi = {10.1103/PhysRevB.102.041119},
  url = {https://link.aps.org/doi/10.1103/PhysRevB.102.041119}
}

@article{Floquet_engineered_tp_5,
  title = {Composite topological phases via Floquet engineering},
  author = {Wu, Hong},
  journal = {Phys. Rev. B},
  volume = {108},
  issue = {19},
  pages = {195125},
  numpages = {6},
  year = {2023},
  month = {Nov},
  publisher = {American Physical Society},
  doi = {10.1103/PhysRevB.108.195125},
  url = {https://link.aps.org/doi/10.1103/PhysRevB.108.195125}
}

@article{Floquet_engineered_tp_6,
  author    = {Wintersperger, Karen and Braun, Christoph and {\"{U}}nal, F. Nur and Eckardt, Andr{\'{e}} and Liberto, Marco Di and Goldman, Nathan and Bloch, Immanuel and Aidelsburger, Monika},
  title     = {Realization of an anomalous Floquet topological system with ultracold atoms},
  journal   = {Nature Physics},
  year      = {2020},
  volume    = {16},
  number    = {10},
  pages     = {1058--1063},
  month     = {oct},
  doi       = {10.1038/s41567-020-0949-y},
  url       = {https://doi.org/10.1038/s41567-020-0949-y}
}

@article{Quantum_phase_tran_1,
  title = {Engineering Floquet dynamical quantum phase transitions},
  author = {Naji, J. and Jafari, R. and Zhou, Longwen and Langari, A.},
  journal = {Phys. Rev. B},
  volume = {106},
  issue = {9},
  pages = {094314},
  numpages = {10},
  year = {2022},
  month = {Sep},
  publisher = {American Physical Society},
  doi = {10.1103/PhysRevB.106.094314},
  url = {https://link.aps.org/doi/10.1103/PhysRevB.106.094314}
}

@article{Quantum_phase_tran_2,
  title = {Dynamical relaxation of correlators in periodically driven integrable quantum systems},
  author = {Aditya, Sreemayee and Samanta, Sutapa and Sen, Arnab and Sengupta, K. and Sen, Diptiman},
  journal = {Phys. Rev. B},
  volume = {105},
  issue = {10},
  pages = {104303},
  numpages = {14},
  year = {2022},
  month = {Mar},
  publisher = {American Physical Society},
  doi = {10.1103/PhysRevB.105.104303},
  url = {https://link.aps.org/doi/10.1103/PhysRevB.105.104303}
}

@article{ultracold_1,
  title = {Coherent Control of Dressed Matter Waves},
  author = {Zenesini, Alessandro and Lignier, Hans and Ciampini, Donatella and Morsch, Oliver and Arimondo, Ennio},
  journal = {Phys. Rev. Lett.},
  volume = {102},
  issue = {10},
  pages = {100403},
  numpages = {4},
  year = {2009},
  month = {Mar},
  publisher = {American Physical Society},
  doi = {10.1103/PhysRevLett.102.100403},
  url = {https://link.aps.org/doi/10.1103/PhysRevLett.102.100403}
}

@article{ultracold_2,
  title = {Experimental Realization of Strong Effective Magnetic Fields in an Optical Lattice},
  author = {Aidelsburger, M. and Atala, M. and Nascimb\`ene, S. and Trotzky, S. and Chen, Y.-A. and Bloch, I.},
  journal = {Phys. Rev. Lett.},
  volume = {107},
  issue = {25},
  pages = {255301},
  numpages = {5},
  year = {2011},
  month = {Dec},
  publisher = {American Physical Society},
  doi = {10.1103/PhysRevLett.107.255301},
  url = {https://link.aps.org/doi/10.1103/PhysRevLett.107.255301}
}

@article{ultracold_3,
  title = {Measuring a localization phase diagram controlled by the interplay of disorder and driving},
  author = {Dotti, Peter and Bai, Yifei and Shimasaki, Toshihiko and Dardia, Anna R. and Weld, David M.},
  journal = {Phys. Rev. Res.},
  volume = {7},
  issue = {2},
  pages = {L022026},
  numpages = {6},
  year = {2025},
  month = {Apr},
  publisher = {American Physical Society},
  doi = {10.1103/PhysRevResearch.7.L022026},
  url = {https://link.aps.org/doi/10.1103/PhysRevResearch.7.L022026}
}

@article{ultracold_4,
  author    = {Goldman, N. and Budich, J. C. and Zoller, P.},
  title     = {Topological quantum matter with ultracold gases in optical lattices},
  journal   = {Nature Physics},
  year      = {2016},
  volume    = {12},
  number    = {7},
  pages     = {639--645},
  month     = {jul},
  doi       = {10.1038/nphys3803},
  url       = {https://doi.org/10.1038/nphys3803}
}

@article{ultracold_5,
  author    = {Kennedy, Colin J. and Burton, William Cody and Chung, Woo Chang and Ketterle, Wolfgang},
  title     = {Observation of Bose–Einstein condensation in a strong synthetic magnetic field},
  journal   = {Nature Physics},
  year      = {2015},
  volume    = {11},
  number    = {10},
  pages     = {859--864},
  month     = {oct},
  doi       = {10.1038/nphys3421},
  url       = {https://doi.org/10.1038/nphys3421}
}

@article{ultracold_6,
  author    = {Tai, M. Eric and Lukin, Alexander and Rispoli, Matthew and Schittko, Robert and Menke, Tim and Borgnia, Dan and Preiss, Philipp M. and Grusdt, Fabian and Kaufman, Adam M. and Greiner, Markus},
  title     = {Microscopy of the interacting Harper–Hofstadter model in the two-body limit},
  journal   = {Nature},
  year      = {2017},
  volume    = {546},
  number    = {7659},
  pages     = {519--523},
  month     = {jun},
  doi       = {10.1038/nature22811},
  url       = {https://doi.org/10.1038/nature22811}
}

@article{ME_in_Flo_1,
  title = {Mobility edge and multifractality in a periodically driven Aubry-Andr\'e model},
  author = {Sarkar, Madhumita and Ghosh, Roopayan and Sen, Arnab and Sengupta, K.},
  journal = {Phys. Rev. B},
  volume = {103},
  issue = {18},
  pages = {184309},
  numpages = {16},
  year = {2021},
  month = {May},
  publisher = {American Physical Society},
  doi = {10.1103/PhysRevB.103.184309},
  url = {https://link.aps.org/doi/10.1103/PhysRevB.103.184309}
}

@article{ME_in_Flo_2,
  title = {Mobility edges and critical regions in a periodically kicked incommensurate optical Raman lattice},
  author = {Wang, Yucheng},
  journal = {Phys. Rev. A},
  volume = {106},
  issue = {5},
  pages = {053312},
  numpages = {9},
  year = {2022},
  month = {Nov},
  publisher = {American Physical Society},
  doi = {10.1103/PhysRevA.106.053312},
  url = {https://link.aps.org/doi/10.1103/PhysRevA.106.053312}
}

@article{Floqute_theory_1,
  title = {Solution of the Schr\"odinger Equation with a Hamiltonian Periodic in Time},
  author = {Shirley, Jon H.},
  journal = {Phys. Rev.},
  volume = {138},
  issue = {4B},
  pages = {B979--B987},
  numpages = {0},
  year = {1965},
  month = {May},
  publisher = {American Physical Society},
  doi = {10.1103/PhysRev.138.B979},
  url = {https://link.aps.org/doi/10.1103/PhysRev.138.B979}
}

@article{Floqute_theory_2,
  title = {Stark Effect in Rapidly Varying Fields},
  author = {Autler, S. H. and Townes, C. H.},
  journal = {Phys. Rev.},
  volume = {100},
  issue = {2},
  pages = {703--722},
  numpages = {0},
  year = {1955},
  month = {Oct},
  publisher = {American Physical Society},
  doi = {10.1103/PhysRev.100.703},
  url = {https://link.aps.org/doi/10.1103/PhysRev.100.703}
}

@article{Floquet_singleparticle1,
  title = {Dynamical Control of Matter-Wave Tunneling in Periodic Potentials},
  author = {Lignier, H. and Sias, C. and Ciampini, D. and Singh, Y. and Zenesini, A. and Morsch, O. and Arimondo, E.},
  journal = {Phys. Rev. Lett.},
  volume = {99},
  issue = {22},
  pages = {220403},
  numpages = {4},
  year = {2007},
  month = {Nov},
  publisher = {American Physical Society},
  doi = {10.1103/PhysRevLett.99.220403},
  url = {https://link.aps.org/doi/10.1103/PhysRevLett.99.220403}
}

@article{Floquet_singleparticle2,
  title = {Periodically driven model with quasiperiodic potential and staggered hopping amplitudes: Engineering of mobility gaps and multifractal states},
  author = {Aditya, Sreemayee and Sengupta, K. and Sen, Diptiman},
  journal = {Phys. Rev. B},
  volume = {107},
  issue = {3},
  pages = {035402},
  numpages = {14},
  year = {2023},
  month = {Jan},
  publisher = {American Physical Society},
  doi = {10.1103/PhysRevB.107.035402},
  url = {https://link.aps.org/doi/10.1103/PhysRevB.107.035402}
}

@article{Floquet_manybody_1,
  title = {Fate of Many-Body Localization Under Periodic Driving},
  author = {Lazarides, Achilleas and Das, Arnab and Moessner, Roderich},
  journal = {Phys. Rev. Lett.},
  volume = {115},
  issue = {3},
  pages = {030402},
  numpages = {5},
  year = {2015},
  month = {Jul},
  publisher = {American Physical Society},
  doi = {10.1103/PhysRevLett.115.030402},
  url = {https://link.aps.org/doi/10.1103/PhysRevLett.115.030402}
}

@article{floquet_manybody2,
  title = {Stark many-body localization under periodic driving},
  author = {Duffin, Christian and Deger, Aydin and Lazarides, Achilleas},
  journal = {Phys. Rev. B},
  volume = {110},
  issue = {13},
  pages = {134205},
  numpages = {6},
  year = {2024},
  month = {Oct},
  publisher = {American Physical Society},
  doi = {10.1103/PhysRevB.110.134205},
  url = {https://link.aps.org/doi/10.1103/PhysRevB.110.134205}
}

@article{ME_mosaic,
  author    = {X. Gao and Z.-H. Zhou and J. Cao and W.-X. Cui and S. Zhang and H.-F. Wang},
  title     = {Floquet Engineering of Mobility Edges in Quasiperiodic Mosaic Lattice},
  journal   = {Advanced Quantum Technologies},
  volume    = {8},
  number    = {11},
  pages     = {e00287},
  year      = {2025},
  doi       = {10.1002/qute.202500287}
}

@misc{gpd,
      title={Floquet mobility edges and transport in a periodically driven generalized Aubry-Andr\'e model}, 
      author={Jayashis Das and Vatsana Tiwari and Manish Kumar and Auditya Sharma},
      year={2026},
      eprint={2604.21992},
      archivePrefix={arXiv},
      primaryClass={cond-mat.dis-nn},
      url={https://arxiv.org/abs/2604.21992}, 
}

@article{ladder_peierls,
  title = {Anderson localization via Peierls phase modulation},
  author = {Goswami, Arpita and Chatterjee, Pallabi and Modak, Ranjan and Sahoo, Shaon},
  journal = {Phys. Rev. B},
  volume = {114},
  issue = {9},
  pages = {094204},
  numpages = {18},
  year = {2026},
  month = {Aug},
  publisher = {American Physical Society},
  doi = {10.1103/8zf7-6y67},
  url = {https://link.aps.org/doi/10.1103/8zf7-6y67}
}

\end{document}